\documentclass[]{aastex631}

\usepackage[T1]{fontenc}

\begin{document}

\title{The 2023 outburst of the \textit{Gaia} alerted EXor Gaia23bab}

\author[0000-0002-3632-1194]{Zs\'ofia Nagy}
\affiliation{Konkoly Observatory, HUN-REN Research Centre for Astronomy and Earth Sciences \\ Konkoly-Thege Mikl\'os \'ut 15-17, H-1121 Budapest, Hungary}
\affiliation{CSFK, MTA Centre of Excellence, Konkoly-Thege Mikl\'os \'ut 15-17, 1121 Budapest, Hungary}
\email{nagy.zsofia@csfk.org}

\author[0000-0001-7157-6275]{\'Agnes K\'osp\'al}
\affiliation{Konkoly Observatory, HUN-REN Research Centre for Astronomy and Earth Sciences \\ Konkoly-Thege Mikl\'os \'ut 15-17, H-1121 Budapest, Hungary}
\affiliation{CSFK, MTA Centre of Excellence, Konkoly-Thege Mikl\'os \'ut 15-17, 1121 Budapest, Hungary}
\affiliation{ELTE E\"otv\"os Lor\'and University, Institute of Physics and Astronomy, P\'azm\'any P\'eter s\'et\'any 1A, Budapest 1117, Hungary}

\author[0000-0001-6015-646X]{P\'eter \'Abrah\'am}
\affiliation{Konkoly Observatory, HUN-REN Research Centre for Astronomy and Earth Sciences \\ Konkoly-Thege Mikl\'os \'ut 15-17, H-1121 Budapest, Hungary}
\affiliation{CSFK, MTA Centre of Excellence, Konkoly-Thege Mikl\'os \'ut 15-17, 1121 Budapest, Hungary}
\affiliation{ELTE E\"otv\"os Lor\'and University, Institute of Physics and Astronomy, P\'azm\'any P\'eter s\'et\'any 1A, Budapest 1117, Hungary}
\affiliation{Department of Astrophysics, University of Vienna, Türkenschanzstr. 17, 1180, Vienna, Austria}

\author[0000-0002-7035-8513]{Teresa Giannini}
\affiliation{INAF-Osservatorio Astronomico di Roma, via di Frascati 33, 00078, Monte Porzio Catone, Italy}

\author[0000-0002-7538-5166]{M\'aria Kun}
\affiliation{Konkoly Observatory, HUN-REN Research Centre for Astronomy and Earth Sciences \\ Konkoly-Thege Mikl\'os \'ut 15-17, H-1121 Budapest, Hungary}
\affiliation{CSFK, MTA Centre of Excellence, Konkoly-Thege Mikl\'os \'ut 15-17, 1121 Budapest, Hungary}

\author[0000-0002-8364-7795]{Manuele Gangi}
\affiliation{ASI, Italian Space Agency, Via del Politecnico snc, 00133 Rome, Italy}
\affiliation{INAF – Osservatorio Astronomico di Roma, Via Frascati 33, 00078 Monte Porzio Catone, Italy}

\author[0000-0002-4283-2185]{Fernando Cruz-S\'aenz de Miera}
\affiliation{Institut de Recherche en Astrophysique et Plan\'etologie, Universit\'e de Toulouse, UT3-PS, OMP, CNRS, 9 av. du Colonel Roche, 31028 Toulouse Cedex 4, France}
\affiliation{Konkoly Observatory, HUN-REN Research Centre for Astronomy and Earth Sciences \\ Konkoly-Thege Mikl\'os \'ut 15-17, H-1121 Budapest, Hungary}
\affiliation{CSFK, MTA Centre of Excellence, Konkoly-Thege Mikl\'os \'ut 15-17, 1121 Budapest, Hungary}

\author[0000-0002-0631-7514]{Michael Kuhn}
\affiliation{Centre for Astrophysics Research, University of Hertfordshire, College Lane, Hatfield, AL10 9AB, UK}

\author[0000-0001-5018-3560]{Micha\l\ Siwak}
\affiliation{Mt. Suhora Astronomical Observatory, University of the National Education Commission, ul. Podchora\.zych 2, 30-084 Krak{\'o}w, Poland}
\affiliation{Konkoly Observatory, HUN-REN Research Centre for Astronomy and Earth Sciences \\ Konkoly-Thege Mikl\'os \'ut 15-17, H-1121 Budapest, Hungary}
\affiliation{CSFK, MTA Centre of Excellence, Konkoly-Thege Mikl\'os \'ut 15-17, 1121 Budapest, Hungary}

\author[0000-0002-3648-433X]{M\'at\'e Szil\'agyi}
\affiliation{Konkoly Observatory, HUN-REN Research Centre for Astronomy and Earth Sciences \\ Konkoly-Thege Mikl\'os \'ut 15-17, H-1121 Budapest, Hungary}
\affiliation{CSFK, MTA Centre of Excellence, Konkoly-Thege Mikl\'os \'ut 15-17, 1121 Budapest, Hungary}

\author[0000-0002-5261-6216]{Eleonora Fiorellino}
\affiliation{Instituto de Astrofísica de Canarias, IAC, Vía Láctea s/n, 38205 La Laguna (S.C.Tenerife), Spain}
\affiliation{Departamento de Astrofísica, Universidad de La Laguna, 38206 La Laguna (S.C.Tenerife), Spain}

;------------------------------------------------

\author[0000-0002-0666-3847]{Simone Antoniucci}
\affiliation{INAF-Osservatorio Astronomico di Roma, via di Frascati 33, 00078, Monte Porzio Catone, Italy}

\author[0000-0002-1892-2180]{Katia Biazzo}
\affiliation{INAF-Osservatorio Astronomico di Roma, via di Frascati 33, 00078, Monte Porzio Catone, Italy}

\author[0000-0002-8585-4544]{Attila B\'odi}
\affiliation{Konkoly Observatory, HUN-REN Research Centre for Astronomy and Earth Sciences \\ Konkoly-Thege Mikl\'os \'ut 15-17, H-1121 Budapest, Hungary}
\affiliation{CSFK, MTA Centre of Excellence, Konkoly-Thege Mikl\'os \'ut 15-17, 1121 Budapest, Hungary}
\affiliation{Department of Astrophysical Sciences, Princeton University, 4 Ivy Lane, Princeton, NJ 08544, USA}

\author[0000-0001-6232-9352]{Zs\'ofia Bora}
\affiliation{Konkoly Observatory, HUN-REN Research Centre for Astronomy and Earth Sciences \\ Konkoly-Thege Mikl\'os \'ut 15-17, H-1121 Budapest, Hungary}
\affiliation{CSFK, MTA Centre of Excellence, Konkoly-Thege Mikl\'os \'ut 15-17, 1121 Budapest, Hungary}
\affiliation{ELTE E\"otv\"os Lor\'and University, Institute of Physics and Astronomy, P\'azm\'any P\'eter s\'et\'any 1A, Budapest 1117, Hungary}

\author[0000-0002-6497-8863]{Borb\'ala Cseh}
\affiliation{Konkoly Observatory, HUN-REN Research Centre for Astronomy and Earth Sciences \\ Konkoly-Thege Mikl\'os \'ut 15-17, H-1121 Budapest, Hungary}
\affiliation{CSFK, MTA Centre of Excellence, Konkoly-Thege Mikl\'os \'ut 15-17, 1121 Budapest, Hungary}
\affiliation{MTA-ELTE Lend{\"u}let "Momentum" Milky Way Research Group, Hungary}

\author[0000-0001-9587-1615]{Marek Dr{\'o}{\.z}d{\.z}}
\affiliation{Mt. Suhora Astronomical Observatory, University of the National Education Commission, ul. Podchora\.zych 2, 30-084 Krak{\'o}w, Poland}

\author{\'Agoston Horti-D\'avid}
\affiliation{ELTE E\"otv\"os Lor\'and University, Institute of Physics and Astronomy, P\'azm\'any P\'eter s\'et\'any 1A, Budapest 1117, Hungary}
\affiliation{Konkoly Observatory, HUN-REN Research Centre for Astronomy and Earth Sciences \\ Konkoly-Thege Mikl\'os \'ut 15-17, H-1121 Budapest, Hungary}
\affiliation{CSFK, MTA Centre of Excellence, Konkoly-Thege Mikl\'os \'ut 15-17, 1121 Budapest, Hungary}

\author[0000-0001-5203-434X]{Andr\'as P\'eter Jo\'o}
\affiliation{Konkoly Observatory, HUN-REN Research Centre for Astronomy and Earth Sciences \\ Konkoly-Thege Mikl\'os \'ut 15-17, H-1121 Budapest, Hungary}
\affiliation{CSFK, MTA Centre of Excellence, Konkoly-Thege Mikl\'os \'ut 15-17, 1121 Budapest, Hungary}
\affiliation{ELTE E\"otv\"os Lor\'and University, Institute of Physics and Astronomy, P\'azm\'any P\'eter s\'et\'any 1A, Budapest 1117, Hungary}

\author[0000-0002-1663-0707]{Csilla Kalup}
\affiliation{Konkoly Observatory, HUN-REN Research Centre for Astronomy and Earth Sciences \\ Konkoly-Thege Mikl\'os \'ut 15-17, H-1121 Budapest, Hungary}
\affiliation{CSFK, MTA Centre of Excellence, Konkoly-Thege Mikl\'os \'ut 15-17, 1121 Budapest, Hungary}
\affiliation{ELTE E\"otv\"os Lor\'and University, Institute of Physics and Astronomy, P\'azm\'any P\'eter s\'et\'any 1A, Budapest 1117, Hungary}

\author{Krzysztof Kotysz}
\affiliation{Astronomical Institute, University of Wroc{\l}aw, ul. M. Kopernika 11, 51-622 Wroc{\l}aw, Poland}

\author{Levente Kriskovics}
\affiliation{Konkoly Observatory, HUN-REN Research Centre for Astronomy and Earth Sciences \\ Konkoly-Thege Mikl\'os \'ut 15-17, H-1121 Budapest, Hungary}
\affiliation{CSFK, MTA Centre of Excellence, Konkoly-Thege Mikl\'os \'ut 15-17, 1121 Budapest, Hungary}

\author[0000-0002-1326-1686]{G\'abor Marton}
\affiliation{Konkoly Observatory, HUN-REN Research Centre for Astronomy and Earth Sciences \\ Konkoly-Thege Mikl\'os \'ut 15-17, H-1121 Budapest, Hungary}
\affiliation{CSFK, MTA Centre of Excellence, Konkoly-Thege Mikl\'os \'ut 15-17, 1121 Budapest, Hungary}

\author[0000-0001-8916-8050]{Przemys{\l}aw J. Miko{\l}ajczyk}
\affiliation{Warsaw University Observatory, Al. Ujazdowskie 4, 00-478 Warsaw, Poland}
\affiliation{Astronomical Institute, University of Wroc{\l}aw, ul. M. Kopernika 11, 51-622 Wroc{\l}aw, Poland}

\author[0000-0002-9190-0113]{Brunella Nisini}
\affiliation{INAF-Osservatorio Astronomico di Roma, via di Frascati 33, 00078, Monte Porzio Catone, Italy}

\author[0000-0001-5449-2467]{Andr\'as P\'al}
\affiliation{Konkoly Observatory, HUN-REN Research Centre for Astronomy and Earth Sciences \\ Konkoly-Thege Mikl\'os \'ut 15-17, H-1121 Budapest, Hungary}
\affiliation{CSFK, MTA Centre of Excellence, Konkoly-Thege Mikl\'os \'ut 15-17, 1121 Budapest, Hungary}

\author[0000-0002-3658-2175]{B\'alint Seli}
\affiliation{Konkoly Observatory, HUN-REN Research Centre for Astronomy and Earth Sciences \\ Konkoly-Thege Mikl\'os \'ut 15-17, H-1121 Budapest, Hungary}
\affiliation{CSFK, MTA Centre of Excellence, Konkoly-Thege Mikl\'os \'ut 15-17, 1121 Budapest, Hungary}
\affiliation{ELTE E\"otv\"os Lor\'and University, Institute of Physics and Astronomy, P\'azm\'any P\'eter s\'et\'any 1A, Budapest 1117, Hungary}

\author[0000-0001-7806-2883]{\'Ad\'am S\'odor}
\affiliation{Konkoly Observatory, HUN-REN Research Centre for Astronomy and Earth Sciences \\ Konkoly-Thege Mikl\'os \'ut 15-17, H-1121 Budapest, Hungary}
\affiliation{CSFK, MTA Centre of Excellence, Konkoly-Thege Mikl\'os \'ut 15-17, 1121 Budapest, Hungary}

\author[0000-0002-2046-4131]{L\'aszl\'o Szabados}
\affiliation{Konkoly Observatory, HUN-REN Research Centre for Astronomy and Earth Sciences \\ Konkoly-Thege Mikl\'os \'ut 15-17, H-1121 Budapest, Hungary}
\affiliation{CSFK, MTA Centre of Excellence, Konkoly-Thege Mikl\'os \'ut 15-17, 1121 Budapest, Hungary}

\author{Norton Oliv\'er Szab\'o}
\affiliation{ELTE E\"otv\"os Lor\'and University, Institute of Physics and Astronomy, P\'azm\'any P\'eter s\'et\'any 1A, Budapest 1117, Hungary}
\affiliation{Konkoly Observatory, HUN-REN Research Centre for Astronomy and Earth Sciences \\ Konkoly-Thege Mikl\'os \'ut 15-17, H-1121 Budapest, Hungary}
\affiliation{CSFK, MTA Centre of Excellence, Konkoly-Thege Mikl\'os \'ut 15-17, 1121 Budapest, Hungary}

\author[0000-0001-9830-3509]{Zs\'ofia Marianna Szab\'o}
\affiliation{Max Planck Institute for Radio Astronomy, Auf dem Hügel 69, 53121 Bonn, Germany}
\affiliation{Scottish Universities Physics Alliance (SUPA), School of Physics and Astronomy, University of St Andrews, North Haugh, St Andrews, KY16 9SS, UK}
\affiliation{Konkoly Observatory, HUN-REN Research Centre for Astronomy and Earth Sciences \\ Konkoly-Thege Mikl\'os \'ut 15-17, H-1121 Budapest, Hungary}
\affiliation{CSFK, MTA Centre of Excellence, Konkoly-Thege Mikl\'os \'ut 15-17, 1121 Budapest, Hungary}

\author[0000-0002-1698-605X]{R\'obert Szak\'ats}
\affiliation{Konkoly Observatory, HUN-REN Research Centre for Astronomy and Earth Sciences \\ Konkoly-Thege Mikl\'os \'ut 15-17, H-1121 Budapest, Hungary}
\affiliation{CSFK, MTA Centre of Excellence, Konkoly-Thege Mikl\'os \'ut 15-17, 1121 Budapest, Hungary}

\author{V\'azsony Varga}
\affiliation{ELTE E\"otv\"os Lor\'and University, Institute of Physics and Astronomy, P\'azm\'any P\'eter s\'et\'any 1A, Budapest 1117, Hungary}
\affiliation{Konkoly Observatory, HUN-REN Research Centre for Astronomy and Earth Sciences \\ Konkoly-Thege Mikl\'os \'ut 15-17, H-1121 Budapest, Hungary}
\affiliation{CSFK, MTA Centre of Excellence, Konkoly-Thege Mikl\'os \'ut 15-17, 1121 Budapest, Hungary}

\author[0000-0001-8764-7832]{J\'ozsef Vink\'o}
\affiliation{Konkoly Observatory, HUN-REN Research Centre for Astronomy and Earth Sciences \\ Konkoly-Thege Mikl\'os \'ut 15-17, H-1121 Budapest, Hungary}
\affiliation{CSFK, MTA Centre of Excellence, Konkoly-Thege Mikl\'os \'ut 15-17, 1121 Budapest, Hungary}
\affiliation{ELTE E\"otv\"os Lor\'and University, Institute of Physics and Astronomy, P\'azm\'any P\'eter s\'et\'any 1A, Budapest 1117, Hungary}

\author[0000-0002-9658-6151]{{\L}ukasz Wyrzykowski}
\affiliation{Warsaw University Observatory, Al. Ujazdowskie 4, 00-478 Warsaw, Poland}

\author[0000-0001-6434-9429]{Pawe{\l} Zieli{\'n}ski}
\affiliation{Institute of Astronomy, Faculty of Physics, Astronomy and Informatics, Nicolaus Copernicus University in Toru{\'n},\\ ul. Grudzi\k{a}dzka 5, 87-100 Toru{\'n}, Poland}



\begin{abstract}
Episodic accretion is a fundamental process in the build-up of the stellar mass. EX Lupi-type eruptive young stars (EXors) represent one of the main types of episodic accretion.
We study the recently discovered EXor Gaia23bab during its 2023 outburst. 
We obtained optical and near-infrared photometry and spectroscopy to probe the variation of the physical properties of Gaia23bab during its recent outburst. We also collected archival photometry to study a previous outburst of the star.
We used several accretion tracers, including the Ca\,{\sc{ii}} triplet, He\,{\sc{i}}, and various hydrogen lines from the Paschen and Brackett series, to measure the accretion rate during the outburst. 
The accretion rate is consistent with $\sim 2.0 \times 10^{-7} M_\odot$ $\rm{yr}^{-1}$.
Comparing the line fluxes of the hydrogen Brackett series to predictions of Case B theory suggests excitation temperatures of 5000 -- 10000~K and electron densities of \textbf{$10^9$--$10^{10}$} cm$^{-3}$. 
Comparison to the predictions of a model for T Tauri stars revealed that the fluxes of the Balmer series are consistent with temperatures of 5000 -- 12500~K and a hydrogen density of $10^8$ cm$^{-3}$, while the fluxes of the Paschen series are consistent with temperatures in the range between 10000 and 12500~K and a hydrogen density of $10^{11}$ cm$^{-3}$.
The derived temperatures and densities confirm that Gaia23bab is a prototypical EXor, not only due to its accretion rate, but also based on the best fit temperatures and densities revealed by the detected hydrogen lines. 
\end{abstract}

\keywords{Eruptive variable stars(476) -- Stellar accretion(1578) -- Pre-main sequence stars(1290)
-- Star formation(1569)}


\section{Introduction}
\label{sec:intro}

Low-amplitude short term photometric variability ($\sim$0.2 mag over hours to weeks) has been detected in $\sim$50\% of young stellar objects (YSOs, \citealp{Megeath2012}). A subclass of YSOs, called young eruptive stars, shows much larger variability amplitudes of 2-5 mag at optical and infrared wavelengths on longer timescales, from months to a century \citep{HartmannKenyon1996,Audard2014}.
Young eruptive stars experience outburst events, during which their luminosity increases by up to two orders of magnitude. The outbursts are caused by a sudden increase of the mass accretion rate from the circumstellar disk onto the stellar surface, rising from $10^{-10}$--$10^{-8}$ $M_\odot$ yr$^{-1}$ in quiescence to $10^{-6}$--$10^{-4}$ $M_\odot$ yr$^{-1}$ in outburst.
Recent works suggest that episodic accretion is a common process of star formation (e.g. \citet{Fischer2023} and references therein).
Young eruptive stars have traditionally been classified into two main types: EX Lupi-type (EXor) objects \citep{Herbig2008} and FU Orionis-type (FUor) objects \citep{Herbig1977}.

Our understanding of the physics of eruptive young stars and their role in the star formation process is still incomplete. There are several open questions, as recently reviewed by \citet{Fischer2023}. What is the role of variable accretion in determining the stellar mass? Do all stars experience episodic accretion? What are the triggering mechanisms of events of episodic accretion? What physical processes lead to the end of outbursts? How can we investigate these processes based on the observed variability? What is the role of episodic accretion in the process of planet formation?
To better investigate these open questions, it is important to identify new such objects as well as monitor the photometric and spectroscopic properties of the known eruptive YSOs.
An opportunity to find new cases of eruptive young stars, as well as monitor the brightness of already identified targets, is provided by the \textit{Gaia} Photometric Science Alerts system \citep{Hodgkin2021} with its 4$\pi$ sky coverage and approximately monthly cadence, and it already resulted in several discoveries.
Gaia17bpi \citep{Hillenbrand2018}, Gaia18dvy \citep{SzegediElek2020}, and Gaia21elv \citep{Nagy2023} were identified as FUors, Gaia18dvz \citep{Hodapp2019}, Gaia20eae \citep{CruzSaenzdeMiera2022, Ghosh2022}, and Gaia19fct \citep{Park2022} were identified as EXors. Some sources showed both FUor and EXor characteristics, such as Gaia19ajj \citep{Hillenbrand2019}, Gaia19bey \citep{Hodapp2020}, Gaia21bty \citep{Siwak2023}, and Gaia18cjb \citep{Fiorellino2024}. 
Another young star, Gaia20fgx, showed brightness variations similar to EXors, but its accretion rate turned out to be lower than the typical values for EXors \citep{Nagy2022}.

Gaia23bab triggered the \textit{Gaia} Alerts system on 2023 March 6 due to its 2 mag brightening. This source was previously known as a YSO candidate \citep{Marton2016}.
The source is also known as SPICY 97589 \citep{Kuhn2021}, and was suggested to be an EXor candidate based on photometric data by \citet{Kuhn2023}. Recently, \citet{Giannini2024} confirmed the EXor nature of the source based on spectroscopy. 
\citet{Kuhn2023} and \citet{Giannini2024} found its spectral index to be consistent with typical values of Class II sources. 
\citet{Kuhn2023} investigated that Gaia23bab is part of the cluster G38.3-0.9, and estimated its distance to be 900$\pm$45 pc based on Gaia DR3 \citep{GaiaCollaboration2023} parallaxes of six cluster members.

In this paper, we continue the analysis of \citet{Giannini2024}. We provide revised stellar parameters. Based on these stellar parameters we calculate the accretion properties at two epochs during the 2023 outburst. We probe the excitation conditions traced by the H\,{\sc{i}} lines at three epochs, and discuss the detections of other species such as Ca\,{\sc{ii}}, He\,{\sc{i}}, and O\,{\sc{i}}.
We discuss the photometric and spectroscopic data used in this paper in Sect. \ref{sect_observations}, present results on revised stellar parameters, color variations, line detections, accretion rates, H\,{\sc{i}} excitation, and physical properties probed by the Ca\,{\sc{ii}}, He\,{\sc{i}}, and O\,{\sc{i}} lines in Sect. \ref{sect_results}, discuss them in Sect. \ref{sect_discussion}, and summarize them in Sect. \ref{sect_summary}.

\section{Observations}   
\label{sect_observations}

\subsection{Photometry}

We obtained optical photometric observations of Gaia23bab with the 80 cm Ritchey-Chr\'etien telescope (RC80) at the Piszk\'estet\H{o} Mountain Station of Konkoly Observatory (Hungary).
The RC80 telescope was equipped with an FLI PL230 CCD camera, 0$\farcs$55 pixel scale, $18\farcm8\times18\farcm8$ field of view (FoV), Johnson $BV$ and Sloan $g'r'i'$ filters.
We typically obtained 3 images in each filter. We first applied CCD reduction including bias, flatfield, and dark current corrections. Then we performed aperture photometry for the science target and several comparison stars in the FoV using an aperture radius of 2\farcs75. 
We selected those comparison stars from the APASS9 catalog \citep{henden2015} that were within 6$\farcm$5 of the target and which were mostly constant, i.e., the rms of their $V$-band observations from the ASAS-SN Photometry Database \citep{shappee2014, jayasinghe2019} were below 0.1 mag. The APASS9 catalog provided Bessell $BV$ and Sloan $g'r'i'$ magnitudes for the comparison stars. We used the comparison stars for the photometric calibration by fitting a linear color term. Magnitudes taken with the same filter on the same night were averaged. The final uncertainties are the quadratic sum of the formal uncertainties of the aperture photometry, the photometric calibration, and the scatter of the individual magnitudes that were averaged per night.
The results can be found in Tables \ref{tab:phot1} and \ref{tab:phot2} in the Appendix.

Observations were performed with the 60\,cm Carl-Zeiss telescope at Mount Suhora Observatory of the Cracow Pedagogical University (Poland). 
The telescope at Mount Suhora was equipped with an Apogee Aspen-47 camera, 1$\farcs$116 pixel scale, $19\farcm0\times19\farcm0$ FoV, and Johnson $BVRI$ filters. After standard reduction steps on bias, dark and flatfield, these observations were uploaded to the BHTOM service\footnote{BHTOM - Black Hole TOM: \texttt{https://bhtom.space}}, where they were converted to standard magnitudes.
The results are shown in Tab.~\ref{tab:phot3} in the Appendix.

We used the NOTCam instrument on the Nordic Optical Telescope on 2023 March 14 and 15 to obtain NIR photometry (Program ID: 66-109; PI: F. Cruz-S\'aenz de Miera). The instrument includes a $1024 \times 1024$ pixel HgCdTe Rockwell Science Center “HAWAII” array, and for wide-ﬁeld (WF) imaging it has a $4' \times 4'$ FoV (pixel scale: 0\farcs234). We obtained 5 images in each of the $JHK_S$ bands with 4 s exposures. The data were reduced using our own IDL routines. Data reduction steps included sky subtraction, ﬂat-ﬁelding, and co-adding exposures by dither position and ﬁlter. To calibrate the photometry, we used the Two Micron All Sky Survey (2MASS) catalog \citep{cutri2003}. The instrumental magnitudes of the target and all good-quality 2MASS stars in the ﬁeld were extracted using an aperture radius of 2$''$ in all filters. We determined a constant offset between the instrumental and the 2MASS magnitudes by averaging typically 20-30 stars.
The resulting magnitudes are $J=13.560\pm0.022$\,mag, $H=12.382\pm0.013$\,mag, $K_s=11.583\pm0.022$\,mag.

We also took $JHK_s$ images of Gaia23bab with the infrared guiding camera of SpeX, a medium-resolution infrared spectrograph \citep{Rayner2003} on the NASA Infrared Telescope Facility (IRTF) on Mauna Kea, Hawai'i (USA) on 2023 September 1 (Program ID: 2023B037; PI: \'A. K\'osp\'al) using a 5-point dither pattern and 5\,s exposures. We used the dithering to make a sky image which we subtracted from the shifted and co-added images. Finally we obtained aperture photometry for Gaia23bab and several other stars in the approximately $1'\times1'$ FoV as comparison stars. We used 2MASS for the photometric calibration. The resulting magnitudes are $J=13.546\pm0.056$\,mag, $H=12.443\pm0.097$\,mag, $K_s=11.593\pm0.095$\,mag.

In addition to our own photometry, we used archival optical and infrared photometry.
We used mid-infrared photometry from the Wide-field Infrared Survey Explorer (\textit{WISE}, \citealp{Wright2010}) and \textit{NEOWISE} \citep{Mainzer2011} surveys from the NASA/IPAC Infrared Science Archive. \textit{NEOWISE} observed the full sky on average twice per year with multiple exposures per epoch. For a comparison with the photometry from other instruments, we computed the average of multiple exposures of each season. 
We derived the average of the uncertainties of the single exposures (err1). We also calculated the standard deviation of the points we averaged per season (err2). For the error of the data points averaged per epoch we used the maximum of err1 and err2.	

We downloaded $G$ band photometry from the \textit{Gaia} Science Alerts Index website. We used $r$ and $g$ band photometry from the Data Release 21 of the Zwicky Transient Facility (ZTF; \citealp{Masci2019}). 
We ignore the data points that have catflag = 32768, as these are likely affected by clouds and/or the Moon.
We also downloaded $o$ (“orange”, 560–820 nm) and $c$ (“cyan”, 420–650 nm) band magnitudes from the Asteroid Terrestrial-impact Last Alert System (ATLAS, \citealp{Tonry2018}, \citealp{Smith2020}, \citealp{Heinze2018}) survey using the ATLAS Forced Photometry web service \citep{Shingles2021}. Data points fainter than 20 mag were not considered in our analysis, given the ATLAS limiting magnitude of $\sim$20 mag \citep{Tonry2018}.
                   
\subsection{Spectroscopy}

We used the NOT/NOTCam to obtain intermediate-resolution ($R = 2500$) $JHK$ spectra on 2023 March 14/15 (Program ID: 66-109; PI: F. Cruz-S\'aenz de Miera). We used the ABBA nodding pattern along the slit. The exposure time was 1176 s in each band. 
The nearby telluric standard star HIP86349 was observed right before the target observation for telluric correction.
The raw data were reduced using IRAF \citep{Tody1986} for sky subtraction, ﬂat-ﬁelding, bad-pixel removal, aperture tracing, and wavelength calibration. Argon and xenon lamp spectra were used for the wavelength calibration. Hydrogen lines in the spectra of the telluric standard were removed, and then the spectra were normalized. The target spectrum was divided by the normalized telluric spectrum to correct for the telluric lines.

We obtained another near-infrared spectrum of Gaia23bab using IRTF/SpeX on 2023 September 1 (Program ID: 2023B037; PI: \'A. K\'osp\'al). We utilized the SXD grating with the $0\farcs8\times15''$ slit, covering the 0.69--2.57$\,\mu$m wavelength range with a spectral resolution of $R = 800$. The total integration time was 2160\,s, split into several exposures to avoid saturation and to enable proper sky subtraction and cosmic hit rejection. For telluric correction, we observed the A0-type star 5 Vul using the same instrument setup and at the same airmass (1.04) as the science target, with a total integration time of 720\,s. We reduced the data using Spextool \citep{Cushing2004,Vacca2003}. Reduction steps included flatfield correction, wavelength calibration using argon  arc spectra, the subtraction of the images taken at the A and B nod positions along the slit for sky subtraction, calculation of a spatial profile along the slit length, the extraction of the 1D spectra in an aperture positioned at the peak of the spatial profile, the combination of the individual exposures, telluric correction and flux calibration, and the merging of the individual orders. Contemporaneous photometric observations were used for flux calibration.

\section{Results}
\label{sect_results}

\subsection{Light and color variations}
\label{sec:light_color}

\begin{figure*}[h]
\centering
\includegraphics[width=15cm, trim={0cm 1.3cm 0cm 0cm},clip]{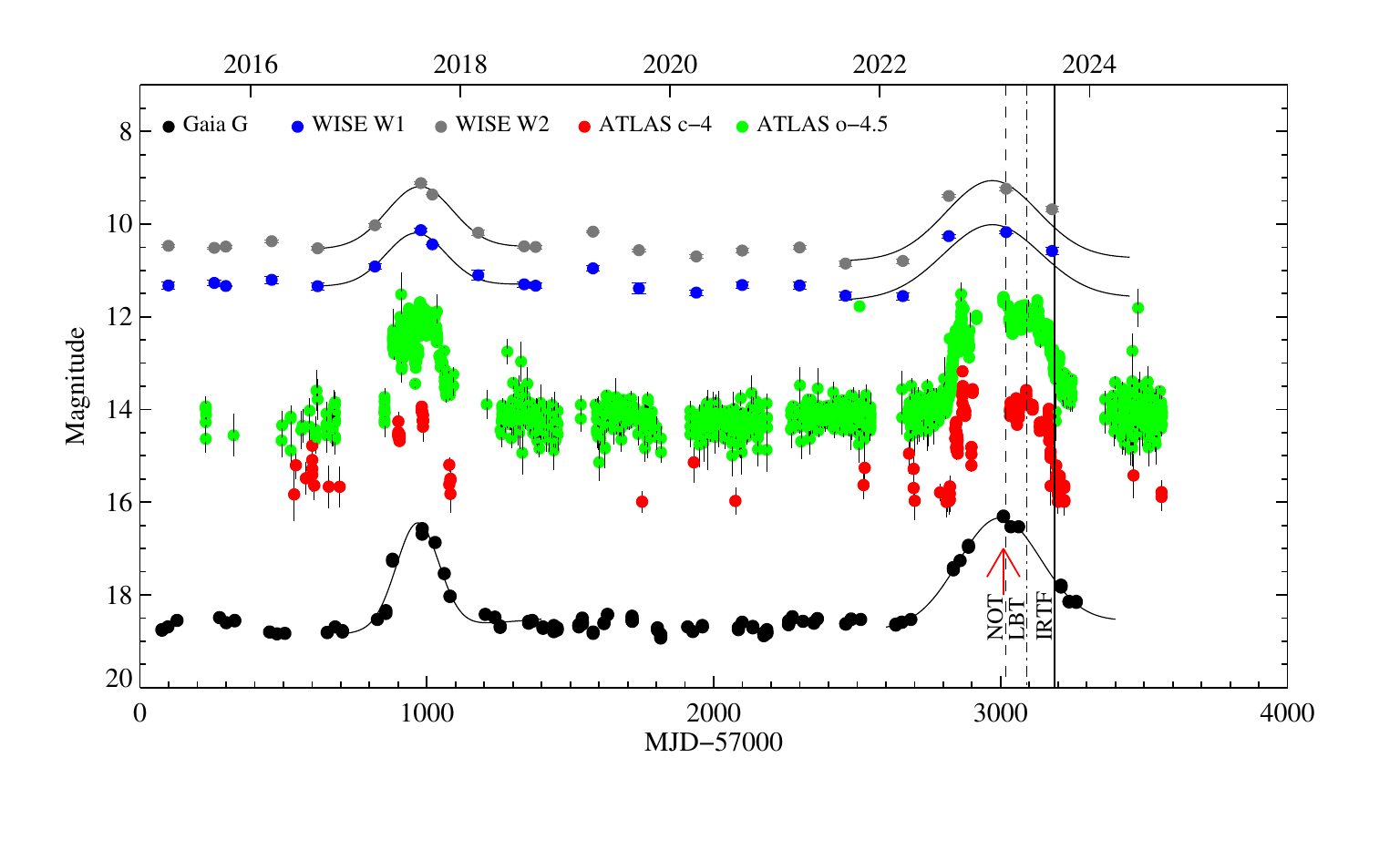}
\includegraphics[width=15cm, trim={0cm 1.3cm 0cm 0cm},clip]{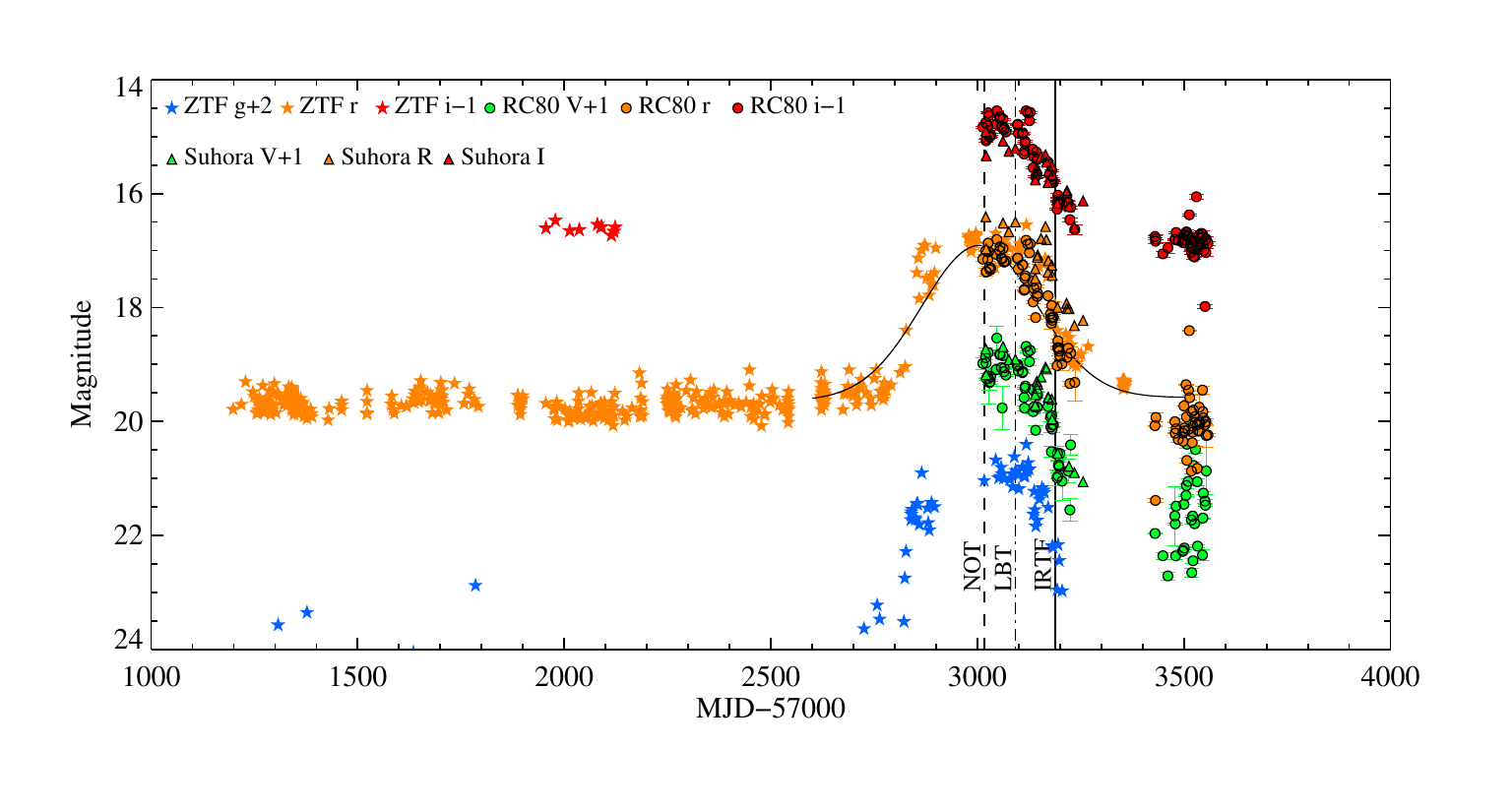}
\caption{\textit{Top panel:} \textit{Gaia} $G$, \textit{WISE}, and ATLAS light curve of Gaia23bab. The red arrow shows the date of the \textit{Gaia} alert. The lines show the epochs of the NOT, LBT, and IRTF spectra.
\textit{Bottom panel:} ZTF, Piszk\'estet\H{o} RC80, and Mt. Suhora light curve of Gaia23bab.
Gaussians fitted to the outbursts are shown in both light curves.
}
\label{fig:lightcurve}
\end{figure*}

The top panel of Figure \ref{fig:lightcurve} shows the \textit{Gaia} $G$, \textit{(NEO)WISE}, and ATLAS light curve of Gaia23bab. In addition to the 2023 outburst, an earlier outburst in 2017 is also seen in the light curve with a duration of about one year (\citealp{Kuhn2023,Giannini2024}). The amplitude ($\sim$2 mag in the \textit{Gaia} $G$-band) is similar for both outbursts. The bottom panel of Fig. \ref{fig:lightcurve} includes the ZTF photometry and the data obtained after the \textit{Gaia} alert using the Piszk\'estet\H{o} RC80 telescope. 
Based on the ATLAS and the Piszk\'estet\H{o} RC80 photometry, Gaia23bab finished its outburst and is back in quiescence. To quantitatively compare the two outbursts, we fitted Gaussians in different bands: in \textit{WISE} $W1$, $W2$, and \textit{Gaia} $G$ for the 2017 outburst, and additionally in ZTF $r$ for the 2023 outburst. The results of the Gaussian fitting is shown in Table \ref{tab:gaussfit} and the fitted Gaussians are shown in Figure \ref{fig:lightcurve}. The duration of the 2023 outburst is longer than that of the 2017 outburst. The amplitude of the outbursts is similar in \textit{Gaia} $G$, but larger for the 2023 outburst in the \textit{WISE} bands. The duration of both outbursts is longer based on the \textit{WISE} light curves than in the optical based on \textit{Gaia} and ZTF. However, the cadence of the \textit{WISE} light curve is lower than that of the optical light curves.

\begin{figure}[h!]
\centering
\includegraphics[width=7cm, trim={0cm 0.5cm 0cm 0.5cm},clip]{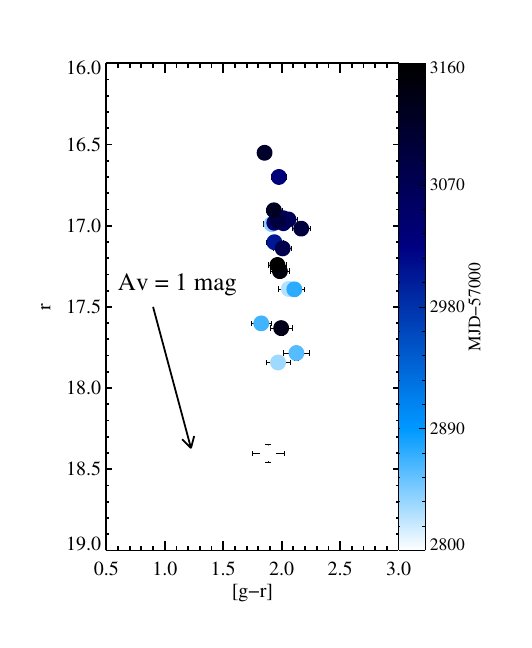}
\includegraphics[width=7cm, trim={0cm 0.5cm 0cm 0.5cm},clip]{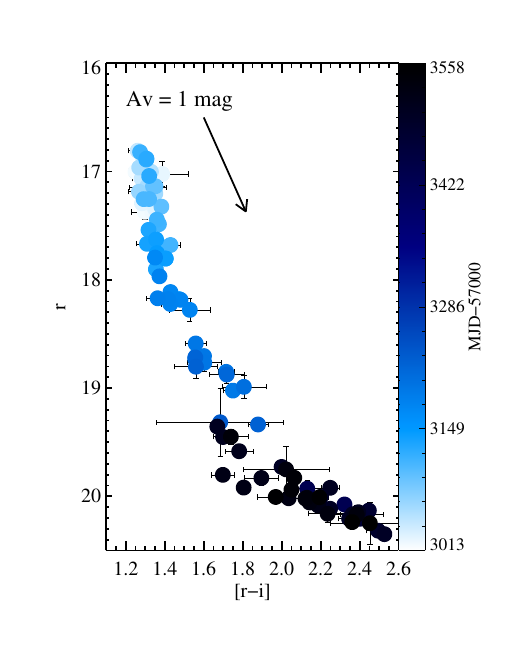}
\includegraphics[width=7cm, trim={0cm 0.5cm 0cm 0.5cm},clip]{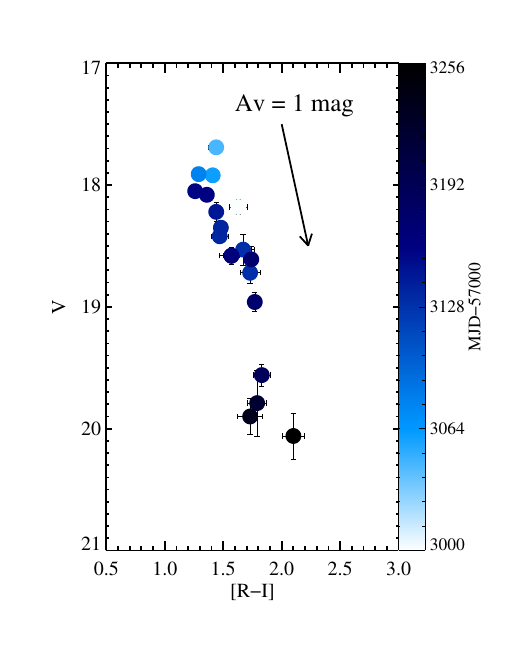}
\includegraphics[width=7cm, trim={0cm 0.5cm 0cm 0.5cm},clip]{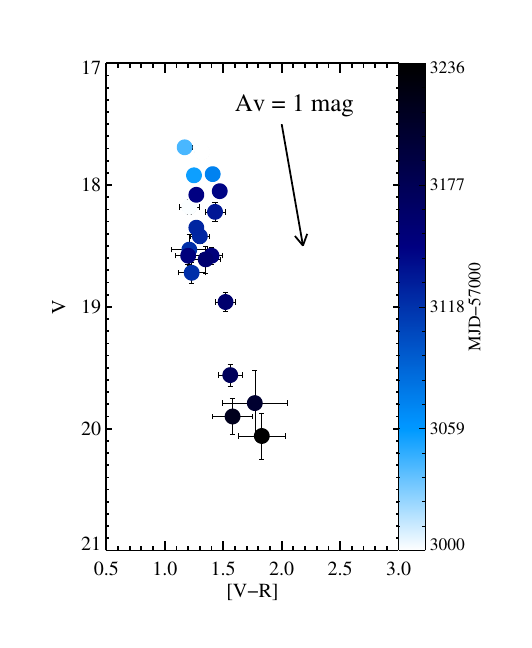}
\caption{
\textit{Top left panel:} Color-magnitude diagram based on ZTF $g$ and $r$ magnitudes mostly during the brightening phase of the 2023 outburst.
\textit{Top right panel:} Color-magnitude diagram during the fading phase of the outburst based on follow-up photometry using the Piszk\'estet\H{o} RC80 telescope.
\textit{Bottom panels:} $V$ vs. $[R-I]$ and $V$ vs. $[V-R]$ color-magnitude diagrams based on Mt. Suhora photometry covering mostly the fading phase.
}
\label{fig:colormag}
\end{figure}

Figure \ref{fig:colormag} shows the $[g-r]$ vs. $g$ color-magnitude diagram based on the ZTF data, which mostly corresponds to the brightening phase of the 2023 outburst. These data show that the color of the brightening is mostly gray. 
The $[r-i]$ vs. $r$ color-magnitude diagram based on photometry obtained with the Piszkestet\H{o} RC80 telescope as well as the $[R-I]$ vs. $V$ and $[V-R]$ vs. $V$ color-magnitude diagrams based on Mt. Suhora data (Fig. \ref{fig:colormag}) during the maximum and the fading phase of the outburst show color variations related to changes in the extinction.

\begin{figure}[h!]
\centering
\includegraphics[width=7cm, trim={0cm 0.5cm 0cm 0.5cm},clip]{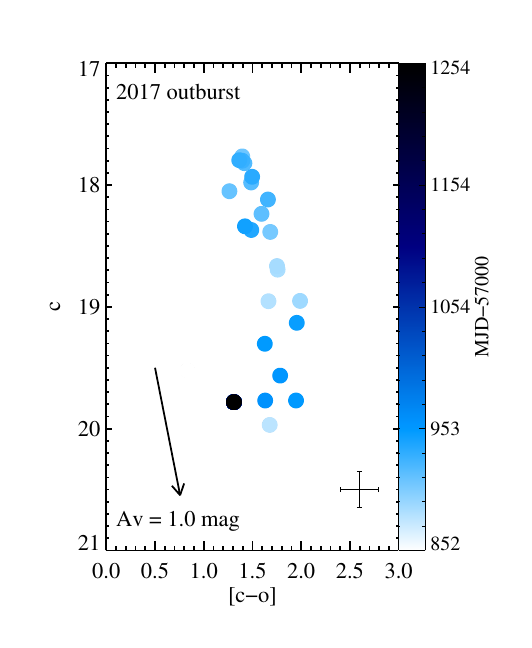}
\includegraphics[width=7cm, trim={0cm 0.5cm 0cm 0.5cm},clip]{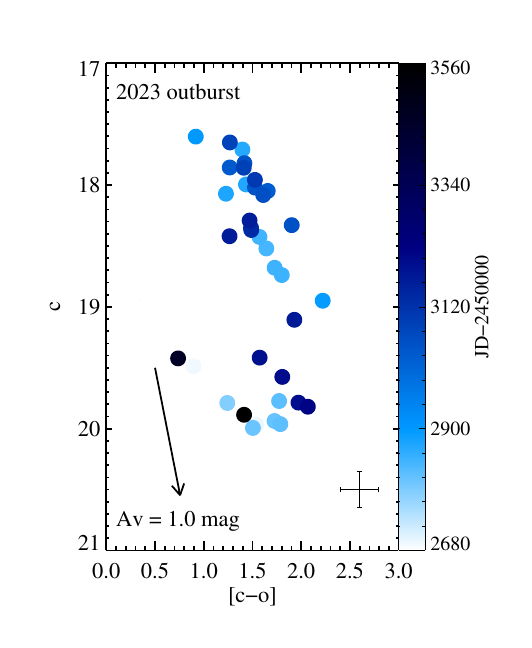}
\caption{
Color-magnitude diagrams during the 2017 (\textit{left panel}) and 2023 (\textit{right panel}) outbursts based on $o$ and $c$ magnitudes from the ATLAS survey. The typical error of the data points is plotted in the lower right corner of the figures.
}
\label{fig:colormag_atlas}
\end{figure}

Photometry from the ATLAS survey is available during both outbursts, which allows a comparison of the color evolution of the 2017 and 2023 outbursts (Fig. \ref{fig:colormag_atlas}). For a comparison of the $o$ and $c$ magnitudes, we interpolated the light curves at the same epochs, and averaged the data points in 3 day bins. As seen in Fig. \ref{fig:colormag_atlas}, both color-magnitude diagrams show color variations which may be partly related to variable circumstellar extinction. Therefore, in addition to the similar amplitude and duration, both outbursts show similar color variations, as was also suggested by $[r-i]$ vs. $[g-r]$ color-color diagrams investigated by \citet{Giannini2024}.  

\begin{figure}[h!]
\centering
\includegraphics[width=7cm, trim={0cm 0.5cm 0cm 0.5cm},clip]{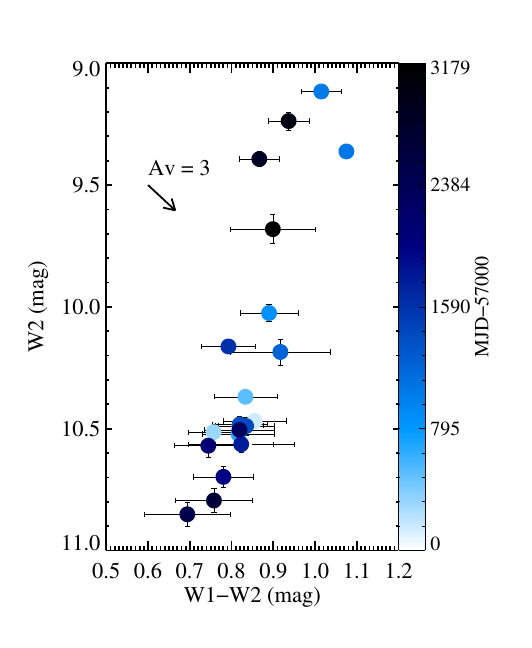}
\caption{Color-magnitude diagram based on \textit{WISE} $W1$ and $W2$ data, covering both the 2017 and 2023 outbursts.}
\label{fig:colormag_wise}
\end{figure}

Figure \ref{fig:colormag_wise} shows a color-magnitude diagram based on the \textit{WISE} $W1$
and $W2$ bands. The source is generally redder when brighter, which is the opposite of what we expect from extinction-related variability, which would cause redder colors when the source is fainter, as also suggested by \citet{Giannini2024}.
Since the $W1$ and $W2$ fluxes originate from the inner regions of the disk, this variability indicates the variations of these disk regions. The areas of the inner disk regions may have increased due to the outburst.

\begin{figure}[h!]
\centering
\includegraphics[width=8.5cm, trim={0cm 0cm 0cm 1cm},clip]{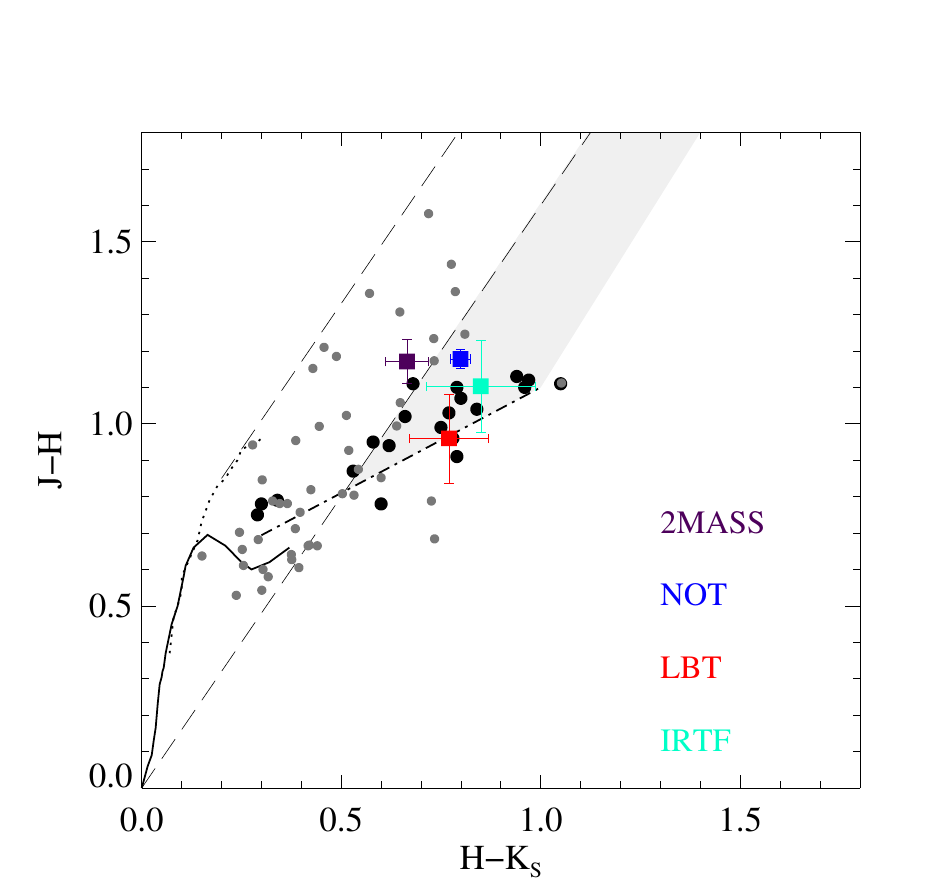}
\caption{($J-H$) versus ($H-K_S$) color–color diagram. 
The solid curve shows the colors of the zero-age main-sequence, and the dotted line
represents the giant branch \citep{BessellBrett1988}. The long-dashed lines
delimit the area occupied by the reddened normal stars \citep{Cardelli1989}.
The dash–dotted line is the locus of unreddened classical T Tauri stars (CTTS) \citep{Meyer1997} and the grey shaded band borders the area of the reddened $K_S$-excess stars.
The black dots correspond to the EXor sample from \citet{Lorenzetti2009}, the smaller grey dots correspond to the young stellar cluster NGC 7023 from \citet{Szilagyi2021}.
}
\label{fig:infra_color}
\end{figure}

Figure \ref{fig:infra_color} shows the [$J-H$] vs. [$H-K_S$] color-color diagrams based on 2MASS photometry, Large Binocular Telescope (LBT) photometry from \citet{Giannini2024} obtained during the 2023 outburst, the data point obtained with the NOT around the brightest state of the 2023 outburst, and the IRTF data point obtained during the fading phase of the 2023 outburst. 
For a comparison, we plotted the [$J-H$] and [$H-K_S$] colors for the EXor sample of \citet{Lorenzetti2009}, except for PV Cep, as its [$J-H$] and [$H-K_S$] values are outside of the plotted range due to very high extinction. 
The [$J-H$] vs. [$H-K_S$] colors of Gaia23bab at the different epochs are consistent with those of the sample of EXors.

\begin{table}
\centering
\caption{Characteristic timescales and amplitudes of the outbursts in different bands obtained using a Gaussian fitting.}
\label{tab:gaussfit}
\begin{tabular}{lccl}
\hline \hline
Outburst, band& Amplitude (mag)& Center (MJD-57000)& FWHM (days)\\
\hline
2017 \textit{WISE} $W1$&  1.1$\pm$0.1&  962$\pm$16&   244$\pm$37\\
2017 \textit{WISE} $W2$&  1.3$\pm$0.1&  976$\pm$12&   267$\pm$24\\
2017 \textit{Gaia} $G$&   2.3$\pm$0.2&  968$\pm$10&   175$\pm$12\\
2023 \textit{WISE} $W1$&  1.6$\pm$0.2&  2970$\pm$50&  400$\pm$60\\
2023 \textit{WISE} $W2$&  1.7$\pm$0.2&  2970$\pm$50&  380$\pm$60\\
2023 \textit{Gaia} $G$&   2.3$\pm$0.1&  2996$\pm$10&  337$\pm$17\\
2023 ZTF $r$&    2.7$\pm$0.3&  3005$\pm$45&  330$\pm$25\\
\hline
\end{tabular}
\end{table}	

\subsection{Line detections}

\begin{figure*}
\centering
\includegraphics[width=15.5cm, trim={0.5cm 0cm 0cm 0cm},clip]{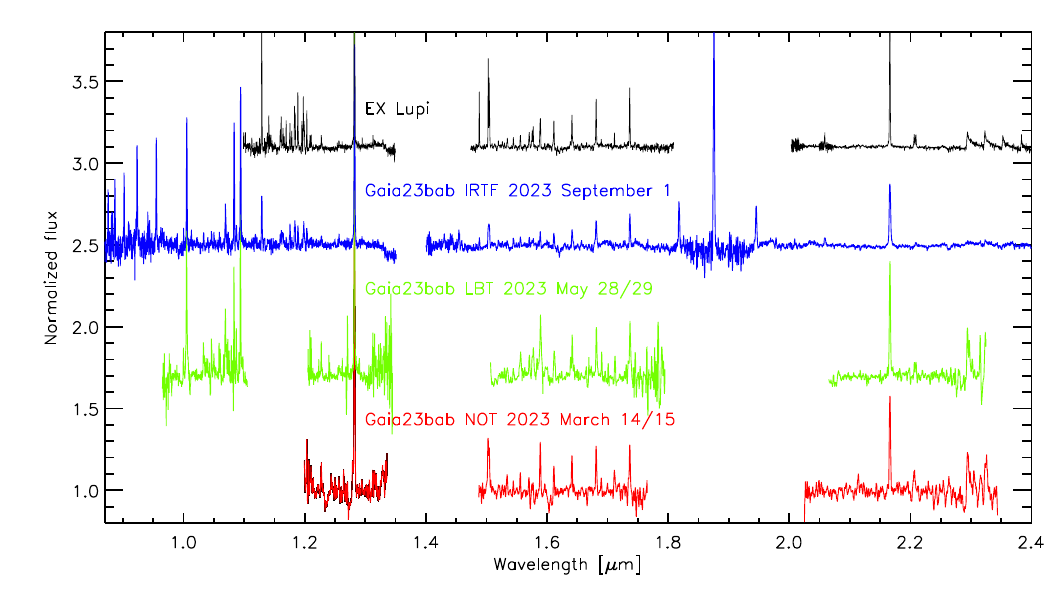}
\caption{Comparison of the NOT, LBT, and IRTF spectra of Gaia23bab with that of EX Lupi from \citep{Kospal2011}.}
\label{fig:spectra_comparison}
\end{figure*}

The first spectrum observed with the NOT was taken on 2023 March 14/15 soon after the \textit{Gaia} alert  was issued on 2023 March 6, close to the maximum brightness of the source. The second spectrum observed on 2023 September 1 with the IRTF was obtained during the fading of the source. Therefore, these spectra represent two different phases of the outburst. For comparison, we include in our analysis the spectrum of \citet{Giannini2024} obtained with the LBT between the NOT and IRTF epochs, on 2023 May 28/29, when the source was already fading. 
Figure \ref{fig:spectra_comparison} shows the NOT, LBT, and IRTF spectra and a spectrum of EX Lupi from \citet{Kospal2011} -- the prototype of EXors -- during outburst. The lines identified in the spectra are shown in Figures \ref{fig:line_ident_1} and \ref{fig:line_ident_2} in Appendix \ref{line_ident}, and their equivalent widths are shown in Tables \ref{tab:spec_lines1} and \ref{tab:spec_lines2}. The profiles of the hydrogen lines are shown in Figures \ref{fig:pa_series} and \ref{fig:br_series} in the Appendix.
The IRTF spectrum has the widest wavelength coverage as seen in Fig. \ref{fig:spectra_comparison}, however, as it was already taken during the fading phase, there are some lines which were detected at the earlier epochs, but not in the IRTF spectrum, such as the CO bandhead. The Na\,{\sc{i}} doublet lines at 2.2 $\mu$m, which are also tracing the disk, are only tentatively detected in the IRTF spectrum, while are clearly seen at the earlier epochs.
Several accretion tracers are detected such as the Ca\,{\sc{ii}} triplet, He\,{\sc{i}} and hydrogen lines from the Paschen series covered by the LBT and IRTF spectra, and from the Brackett series detected at each epoch from Br20 to Br$\gamma$, with the exception of the Br9 and Br$\delta$ lines which are only covered by the IRTF spectrum.
Several metallic lines were detected at each epoch, such as those of Na, Mg, Fe, Si, Ca, Al, C, and O.
The [Fe\,{\sc{ii}}] line at 1.644 $\mu$m, a shock excited line, is detected at each epoch, however, it is blended with the Br12 line. \citet{Giannini2024} reports the weak detection of another shock tracer, the H$_2$ line at 2.1218 $\mu$m, but it is not detected at the other two epochs.

\subsection{Revised stellar parameters}
\label{stellar_param}

\begin{figure*}
\centering
\includegraphics[width=14cm]{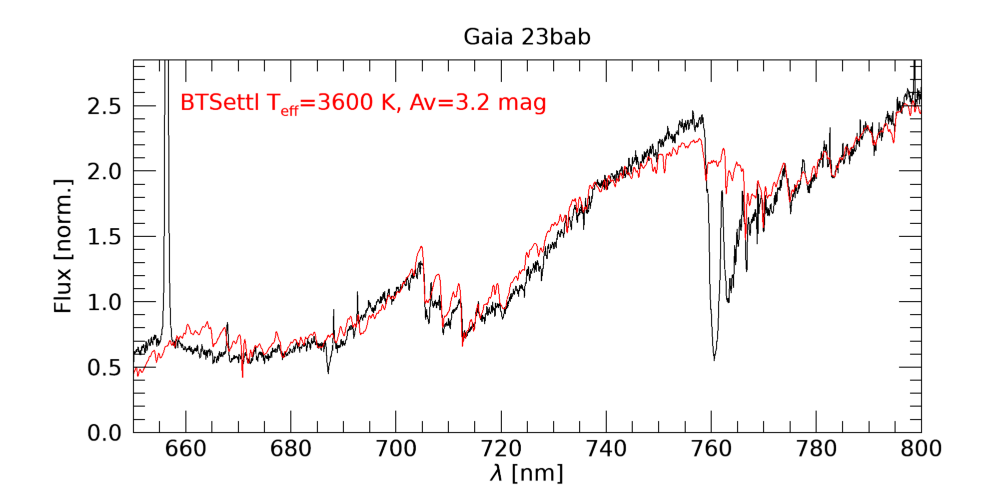}
\caption{Part of the Gaia23bab spectrum obtained with the LBT in quiescence and the best fitting photospheric template.}
\label{fig:spectral_type_fit}
\end{figure*}

In \citet{Giannini2024} an estimate of the stellar parameters was obtained based on the extinction and near-infrared 2MASS photometry in the quiescent phase. 
On 2024 April 11, a quiescence spectrum of Gaia23bab was obtained with the two spectrographs MODS (optical) and LUCI (near-infrared) mounted on the 8.4m Large Binocular Telescope (Arizona, USA). This spectrum will be presented in a forthcoming paper, and here we report the analysis of the optical continuum used to determine the spectral type of Gaia23bab (Figure \ref{fig:spectral_type_fit}). 
The spectrum presents the TiO molecular bands at $\sim$720 nm which can be fitted with an M1 stellar template. In addition, the spectral slope is fitted by assuming A$_V=3.2\pm0.5$ mag.
Knowing the spectral type, we can then estimate the stellar luminosity from the observed magnitudes corrected for the extinction and assuming a bolometric correction (BC). We used the value BC$_J$=1.74 in the $J$ band estimated for $5-30$ Myr stars by \citet{PecautMamajek2013}. From the same paper, we also took the effective temperature ($T_{\rm{eff}}=3630$ K) corresponding to M1 stars.
The stellar luminosity ($L_\star$) is then computed as log($L_\star/L_\odot$)=$0.4(M_{{\rm{bol}},\odot}-M_{\rm{bol}})$, where $M_{{\rm{bol}},\odot}=4.74$ \citep{Mamajek2015} and $M_{\rm{bol}}$ is the intrinsic bolometric magnitude, $M_{\rm{bol}} = m_J -5 \log(d/10({\rm{pc}}))+{\rm{BC}}_J=5.08$.
This results in $L_\star=0.72 \pm 0.07$ $L_\odot$. 
Then, the stellar radius is derived from $L_\star$ and $T_{\rm{eff}}$, assuming that the source emits as a black-body. We estimate $R_\star=2.3\pm0.2~R_\odot$.
Finally, mass is derived from the evolutionary tracks of \citet{Siess2000} as $M_\star=0.40\pm0.05~M_\odot$.

\subsection{Accretion parameters}

\begin{figure*}
\centering
\includegraphics[width=15cm, trim={0.5cm 0.5cm 0cm 0cm},clip]{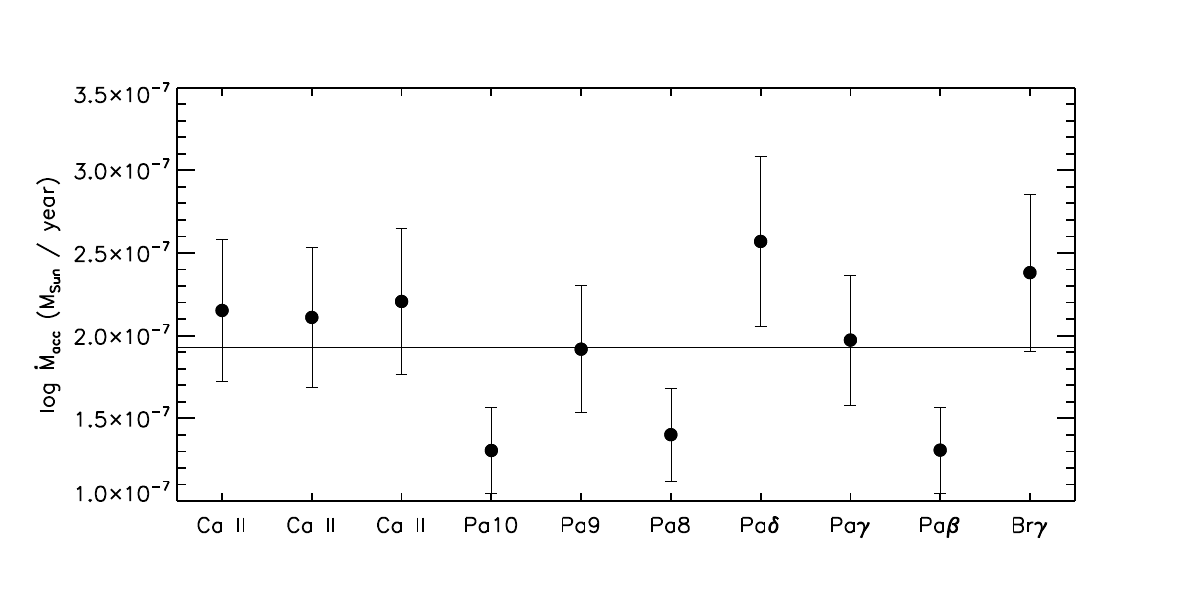}
\caption{The accretion rates derived based on the accretion tracers detected in the IRTF spectrum when assuming an $A_V$ of 3.6 mag and stellar parameters derived in Sect. \ref{stellar_param}.}
\label{fig:acc_rate}
\end{figure*}

In the following section, we derive the accretion rates based on several accretion tracers observed in the NOT and IRTF spectra. 
The line fluxes shown in Table \ref{tab:flux_lines} can be converted to line luminosities after extinction correction as $L_{\rm{line}}=4 \pi d^2 f_{\rm{line}}$, where $d$ is the distance to the source, and $f_{\rm{line}}$ is the extinction-corrected flux of the lines.
We assume the distance of $900\pm45$ pc estimated by \citet{Kuhn2023}.
For the IRTF epoch, we considered different $A_V$ values, as the accretion rates derived from the different lines are supposed to be the same for the right $A_V$. There are only two accretion tracers at the NOT epoch, therefore we assume an $A_V$ of $3.2\pm0.5$ mag for this epoch, as was derived in Sect. \ref{stellar_param}.
We convert the line luminosities to accretion luminosities using the relations derived by \citet{Alcala2017}. 
The accretion luminosities can then be converted to accretion rates using the formula
\begin{equation}
\nonumber
\dot{M}_{\rm{acc}} = 1.25 \frac{L_{\rm{acc}} R_\star}{G M_\star} 
\end{equation}
for which an inner-disk radius of 5 $R_\star$ was assumed \citep{Hartmann1998}.
We assume the stellar radius and mass derived in Sect. \ref{stellar_param}. 
We derive an accretion luminosity of 0.9$\pm$0.2 $L_\odot$ and an accretion rate of $(2.0\pm0.5)\times10^{-7}$ $M_\odot$ yr$^{-1}$ for a best fitting $A_V$ of 3.6 mag at the epoch of the IRTF spectrum. This value is the average value based on all the accretion tracers observed in the IRTF spectrum as shown in Fig. \ref{fig:acc_rate}, and its error is the standard deviation.
At the epoch of the NOT spectrum, only the Pa$\beta$ and Br$\gamma$ lines are available to measure the accretion luminosity and rate, 
which are $\sim$0.8 $L_\odot$ and $\sim$1.4 $L_\odot$ and $1.9\sim10^{-7}$ $M_\odot$ yr$^{-1}$ and $3.2\sim10^{-7}$ $M_\odot$ yr$^{-1}$, respectively.
The accretion rates estimated at the NOT and IRTF epochs are similar to the value of $(2.5\pm0.6)\times10^{-7}$ $M_\odot$ yr$^{-1}$ derived by \citet{Giannini2024} for the epoch of the LBT spectrum, and imply that the accretion rate did not change significantly during the outburst.

\begin{table}
\centering
\caption{Fluxes of the lines (not corrected for extinction) which are used to estimate the accretion parameters from the NOT and IRTF spectra.}
\label{tab:flux_lines}
\begin{tabular}{lcrrr}
\hline \hline
Species& $\lambda_{\rm{tab}}$& $F$(NOT)& $F$(IRTF)\\
       &             ($\mu$m)& \multicolumn{2}{c}{($10^{-15}$ erg s$^{-1}$ cm$^{-2}$)}\\
\hline
Ca\,{\sc{ii}}&            0.8498&         --&              33.6$\pm$5.0\\
Ca\,{\sc{ii}}&            0.8542&         --&              37.8$\pm$5.7\\
Ca\,{\sc{ii}}&            0.8662&         --&              35.0$\pm$5.3\\
H\,{\sc{i}} (Pa10)&       0.9015&         --&               3.8$\pm$0.6\\
H\,{\sc{i}} (Pa9)&        0.9229&          --&              5.5$\pm$0.8\\
H\,{\sc{i}} (Pa8)&        0.9546&         --&               7.3$\pm$1.1\\
H\,{\sc{i}} (Pa$\delta$)& 1.0049&         --&              10.2$\pm$1.5\\
He\,{\sc{i}}&             1.0830&         --&              10.7$\pm$1.6\\
H\,{\sc{i}} (Pa$\gamma$)& 1.0938&         --&              14.7$\pm$2.2\\
H\,{\sc{i}} (Pa$\beta$)&  1.2821&         36.0$\pm$5.4&    24.5$\pm$3.7\\
H\,{\sc{i}} (Br$\gamma$)& 2.1661&         15.4$\pm$2.3&    12.0$\pm$1.8\\
\hline
\end{tabular}
\end{table}	

\subsection{Hydrogen lines}

Strengths of the hydrogen lines provide information about the physical conditions -- temperatures and densities -- of the emitting medium. In the following, we compare the observed Brackett, Paschen, and Balmer decrements to model predictions. We use extinction-corrected fluxes, for which we assume an $A_V$ of 3.6$\pm$0.4 mag, which is based on the accretion rate estimate at the IRTF epoch and the assumption that the $A_V$ did not change significantly between the epochs, which is consistent with the result of the fitting by a stellar template in Sect. \ref{stellar_param}.

\subsubsection{Brackett decrements}

\begin{figure}
\centering
\includegraphics[width=9cm]{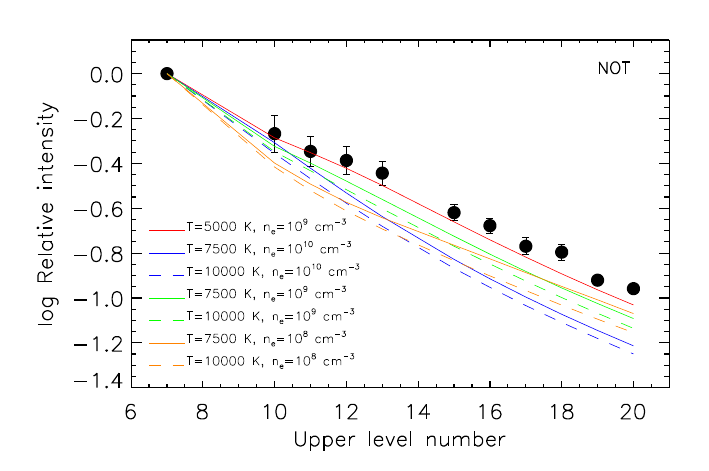} \\
\includegraphics[width=9cm]{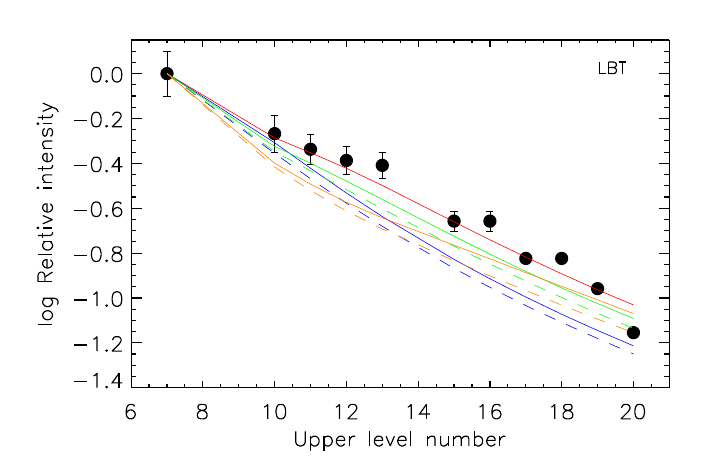} \\
\includegraphics[width=9cm]{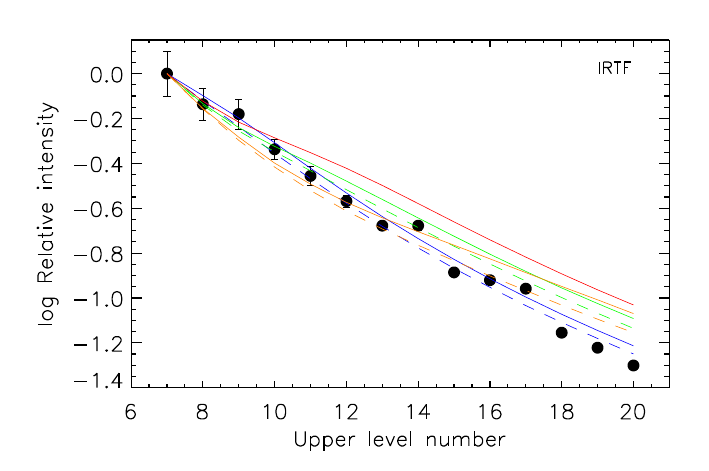}
\caption{
Excitation diagrams for the hydrogen Brackett series for the NOT, LBT, and IRTF spectra. The fluxes have been normalized by the flux of the Br$\gamma$ line. 
The red line shows the best fitting model at the NOT and LBT epochs, and the blue lines show the best fitting models at the IRTF epoch.
}
\label{fig:caseb_brackett}
\end{figure}

To estimate excitation conditions based on the line strengths of the Brackett series, we compare the observed values to the predictions of Case B theory \citep{HummerStorey1987}.
Case B theory assumes that the emitting plasma is opaque to Ly$\alpha$ photons and optically thin for higher transition lines. 
The \citet{HummerStorey1987} models are available for electron densities of 10$^2$, 10$^3$, 10$^4$, 10$^5$, 10$^6$, 10$^7$, 10$^8$, 10$^9$, and 10$^{10}$ cm$^{-3}$ and for temperatures of 1000, 3000, 5000, 7500, 10000, and 12500~K. To find the best fitting model(s) at the three epochs, we applied $\chi^2$ minimization.
The results for the three epochs are shown in Fig. \ref{fig:caseb_brackett}. 
The best fit temperature is 5000~K and the best fit electron density is 10$^9$ cm$^{-3}$ at the NOT and LBT epochs. 
The flux ratios at the IRTF epoch can be better fitted with an electron density of 10$^{10}$ cm$^{-3}$, and are consistent with temperatures of 7500~K and 10000 K.
The best fit parameters at the NOT and LBT epochs are similar to those found using the same method for the eruptive young star V899 Mon \citep{Park2021}. 
The results for each epoch for Gaia23bab differ from the results of \citet{Kospal2011} for EX Lupi, who found a best fit temperature of 10000~K and an electron density of $10^7$ cm$^{-3}$.
However, Case B theory has been applied to investigate the excitation conditions of several other young stellar objects (e.g. \citealp{Bary2008}, \citealp{Nisini2004}, \citealp{Podio2008}, \citealp{VaccaSandell2011}, \citealp{Kraus2012}; \citealp{Whelan2014}), and the estimated temperatures and densities both span a huge range, with values of $T = 1000-20,000$~K and electron densities in the range between $10^7$ and $10^{13}$ cm$^{-3}$. Due to the large range of best fit parameters, and the assumptions of the method, the validity of Case B theory for circumstellar environments of T Tauri stars has been questioned \citep{Edwards2013}.

\subsubsection{Balmer and Paschen decrements}

Local line excitation calculations have been developed by \citet{KwanFischer2011} for conditions appropriate
for winds and accretion flows of T Tauri stars. These calculations give predictions for the Balmer and Paschen series. 
Predictions for the Balmer decrements (ratios from H15 to H$\alpha$ with respect to H$\beta$) in the model are given for densities 8 $<$ log(n$_H$) (cm$^{-3}$) $<$ 12.4 and temperatures between 3750 and 15,000~K.
Predictions for the Paschen decrements by \citet{Edwards2013} (ratios from Pa$\gamma$ to Pa12 with respect to Pa$\beta$) are available for a range of densities (8 $<$ log(n$_H$) (cm$^{-3}$) $<$ 12.4) and temperatures (5000-20,000 K).
The Paschen decrements for Gaia23bab are covered by the LBT and IRTF spectra, while the Balmer decrement is covered by the LBT spectrum. 

\begin{figure}
\centering
\includegraphics[width=9.5cm]{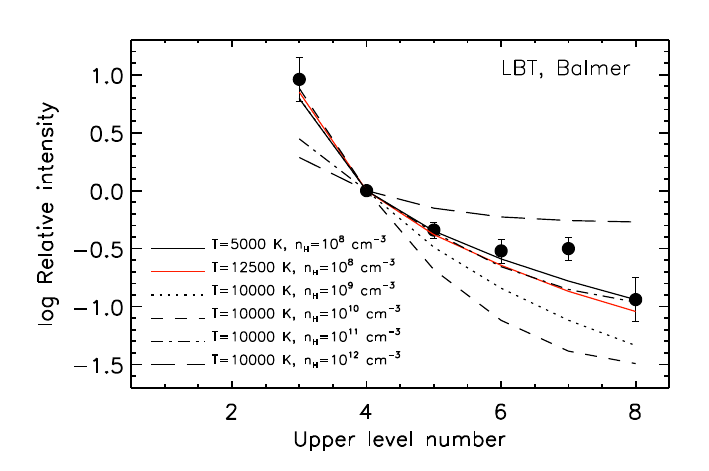} \\
\includegraphics[width=9.5cm]{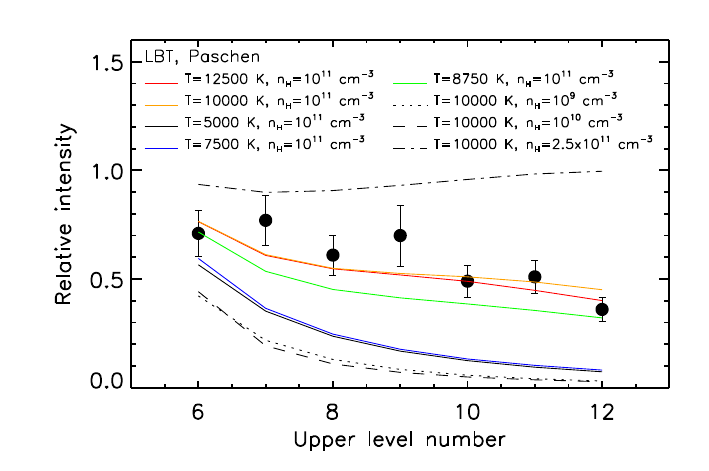} \\
\includegraphics[width=9.5cm]{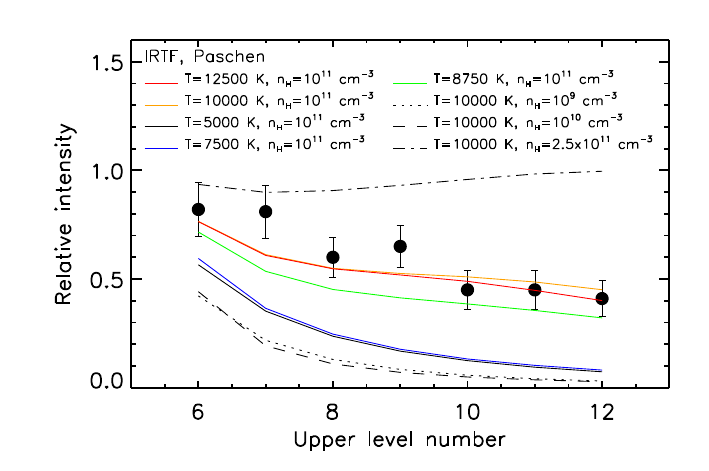}
\caption{Excitation diagrams for the hydrogen Balmer series normalized to the flux of the H$\beta$ line based on the LBT spectrum (top), and the Paschen series normalized to the flux of the Pa$\beta$ line based on the LBT (middle) and IRTF during the outburst (bottom) spectra.  
The overplotted lines correspond to model predictions from \citet{KwanFischer2011} and \citet{Edwards2013}.
}
\label{fig:balmer_paschen_decrement}
\end{figure}

Fluxes of the Balmer series are most consistent with a density of n$_H$ = $10^{8}$ cm$^{-3}$. 
For this density, models with temperatures between 3750~K and 12500~K can reproduce the observed flux ratios. A temperature of 3750~K is unlikely since it is below the stellar temperature (see Sect. \ref{stellar_param}), therefore, the range of temperatures consistent with the Balmer decrement is likely to be between 5000~K and 12500~K. 
Figure \ref{fig:balmer_paschen_decrement} shows models with n$_H$ = $10^{8}$ cm$^{-3}$ and 5000~K as well as 12500~K. It also demonstrates, that the observed flux ratios cannot be fitted with any other H density.

At both epochs, the fluxes of the Paschen series are closest to the predictions of a model with $T$=10000~K and n$_H$ = $10^{11}$ cm$^{-3}$ as well as to a model with $T$=12500~K and n$_H$ = $10^{11}$ cm$^{-3}$ (Fig. \ref{fig:balmer_paschen_decrement}). As a comparison, we plotted other models with different temperatures and densities. A model with the same temperature (10000~K) and a density of $10^{12}$ cm$^{-3}$ is outside of the plotted ranges.

The large range of temperatures found to be consistent with the fluxes of the Balmer series for Gaia23bab is similar to those derived for the sample of EXors studied by \citet{Giannini2022}, however, the best fit density is lower than those found for that EXor sample.
The best fit density of n$_H$ = $10^{11}$ cm$^{-3}$ found for the Paschen series for Gaia23bab was only found in a few cases in the sample of \citep{Giannini2022}, such as PV Cep, DR Tau, and iPTF15afq.
The densities and temperatures found from the Balmer and Paschen decrements for Gaia23bab are similar to those measured for other EXors. The best fit temperatures are consistent with the range that can be expected for H\,{\sc{i}} emission in accretion columns ($T = 6000-20,000$ K; e.g. \citealp{Martin1996}; \citealp{Muzerolle2001}).

\subsection{Ca\,{\sc{ii}}, He\,{\sc{i}}, and O\,{\sc{i}} lines}

The Ca\,{\sc{ii}} triplet, the He\,{\sc{i}}, and O\,{\sc{i}} lines are also related to the accretion process, but may trace different regions of the accretion flow compared to the H\,{\sc{i}} lines. Certain ratios of these lines may include additional information on excitation conditions in Gaia23bab. Based on the data presented here, we analyse the Pa$\gamma$/10,830 He\,{\sc{i}} ratio as well as the Ca\,{\sc{ii}} 8498 / O\,{\sc{i}} 8446 and the Ca\,{\sc{ii}} 8498 / Pa$\gamma$ ratios and compare them to predictions of the \citet{KwanFischer2011} models.

The 10,830 He\,{\sc{i}} line which is observed in the LBT and IRTF spectra is the only allowed radiative transition following the 5876 He\,{\sc{i}} transition. The comparison of the 10,830 He\,{\sc{i}} line to Pa$\gamma$ is straightforward due to the proximity of their wavelengths, as well as due to their density dependence shown in \citet{KwanFischer2011} in their Figure 8.
The Pa$\gamma$/He\,{\sc{i}} ratio is $\sim$1.9 at the LBT epoch and $\sim$1.4 at the IRTF epoch. 
Based on Figure 8 of \citet{KwanFischer2011}, these line ratios are mostly consistent with H densities between $10^{10}$ and $10^{11}$ cm$^{-3}$.
The change in the Pa$\gamma$/He\,{\sc{i}} ratio between the LBT and IRTF epochs could result from decreased optical depth (see Figure 8 in \citealp{KwanFischer2011}).
Pa$\gamma$/He\,{\sc{i}} 10830 ratios $>$1 are not typical of CTTS (e.g., Figure 10 in \citealp{KwanFischer2011}). The Pa$\gamma$/He\,{\sc{i}} 10830 ratio was also found to be below 1 for another EXor, V1741 Sgr \citep{kuhn2024}.

We can compare the Ca\,{\sc{ii}} triplet lines to Pa$\gamma$, as it is a strong hydrogen line not far in wavelength, and to O\,{\sc{i}} 8446 due to its proximity in wavelength as well as its relation to hydrogen excitation via Ly$\beta$ fluorescence.
The Ca\,{\sc{ii}} 8498 / O\,{\sc{i}} 8446 as well as the Ca\,{\sc{ii}} 8498 / Pa$\gamma$ line ratios have similar dependencies on $N$(H) and $T$ and have been found to belong to two main groups \citep{KwanFischer2011}: for one of these groups the Ca\,{\sc{ii}} 8498 / O\,{\sc{i}} 8446 and Ca\,{\sc{ii}} 8498 / Pa$\gamma$ line ratios are close to or less than 1, and for the other group they are larger than 5. For Gaia23bab the Ca\,{\sc{ii}} 8498 / O\,{\sc{i}} 8446 line ratios are $\sim$14.3 and $\sim$12.4, and the Ca\,{\sc{ii}} 8498 / Pa$\gamma$ line ratios are $\sim$3.7 and $\sim$4.5 for the LBT and the IRTF epochs, respectively. Therefore, they are mostly consistent with the second group mentioned by \citet{KwanFischer2011}, and require H densities of $\sim$10$^{12}$ cm$^{-3}$ and temperatures $\lesssim$7500 K. This higher density component compared to the results from the hydrogen lines may be related to the disk boundary layer, as Ca\,{\sc{ii}} line emission may also be related to that in addition to the accretion column \citep{KwanFischer2011}.
In addition, the ratios of the Ca\,{\sc{ii}} triplet lines are close to 1 at both epochs, which implies that they are optically thick. According to the models of \citet{Azevedo2006}, the 0.8542 $\mu$m line is the strongest, which is consistent with the observations for Gaia23bab.

\section{Discussion}
\label{sect_discussion}

\begin{figure}
\centering
\includegraphics[width=7.5cm, trim={0.5cm 0cm 0cm 1.5cm},clip]{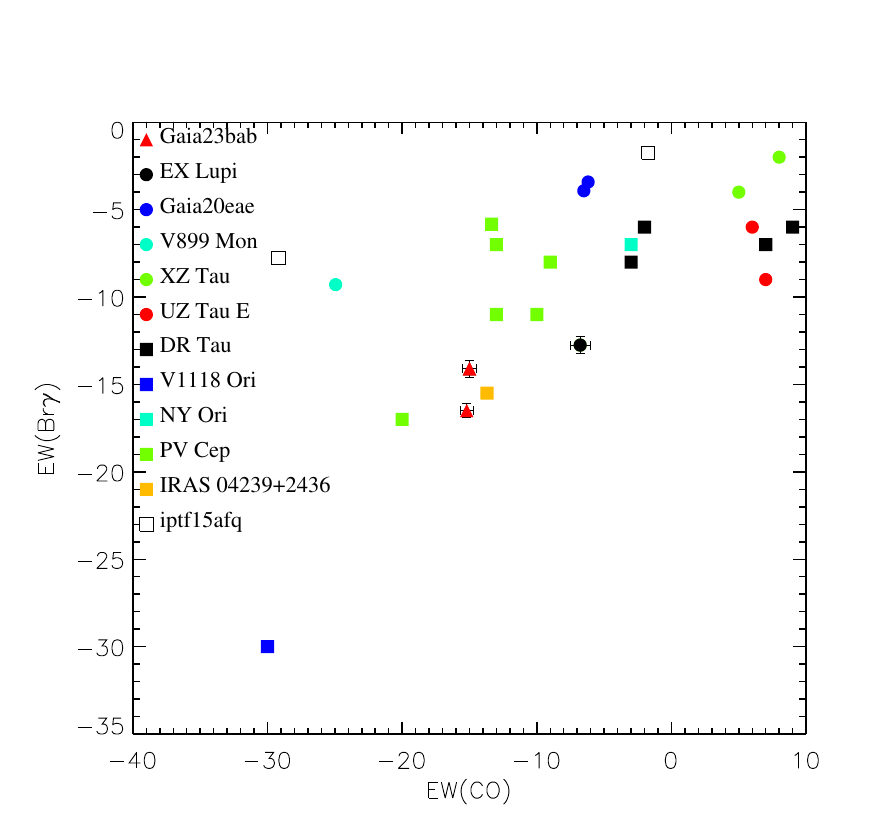}
\includegraphics[width=7.5cm, trim={0.5cm 0cm 0cm 1.5cm},clip]{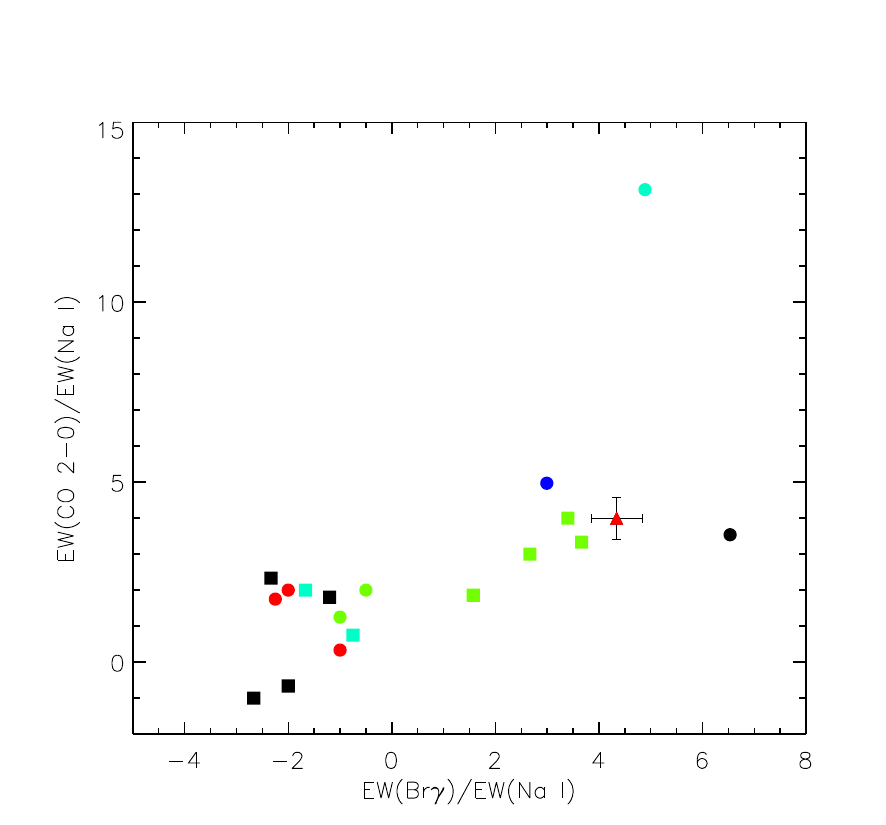}
\caption{Comparison between the equivalent widths of the Br$\gamma$ line and the CO 2-0 line for EXors and other eruptive YSOs with an emission line spectrum. References: EX Lupi \citep{Kospal2011}, Gaia20eae \citep{CruzSaenzdeMiera2022}, V899 Mon \citep{Park2021}, XZ Tau, UZ Tau E, DR Tau, V1118 Ori, NY Ori, PV Cep \citep{Lorenzetti2009}, IRAS 04239+2436 \citep{GreeneLada1996}.}
\label{fig:ew_comparison_co_brgamma}
\end{figure}

Here we compare the properties of Gaia23bab to those of other EXors and eruptive YSOs. The wavelengths of the Br$\gamma$ and CO 2-0 lines are close enough that differences in veiling and extinction can be ignored.
These diagnostics were previously used for other (eruptive) young stars (\citealp{ConnelleyGreene2010}, \citealp{Park2021}).
In Fig. \ref{fig:ew_comparison_co_brgamma} we compare the equivalent widths of the Br$\gamma$ and CO 2-0 lines measured for Gaia23bab and other EXors. Another peculiar eruptive YSO, which represents an intermediate case between EXors and FUors, with an emission line spectrum (V899 Mon, \citealp{Park2021}) is also included, as well as the Class I type YSO IRAS 04239+2436 \citep{GreeneLada1996}.
The general trend of increasing CO emission for increasing Br$\gamma$ emission, which was already seen for other young stars (e.g. \citealp{ConnelleyGreene2010}), is also seen for the plotted EXors. Among the plotted examples, the equivalent widths of Gaia23bab are closest to those of IRAS 04239+2436 \citep{GreeneLada1996} and to PV Cep at some of the observed epochs \citep{Lorenzetti2009}.

Another key diagnostic of EXors is the Na\,{\sc{i}} feature at 2.206 $\mu$m. Given that it is close to the Br$\gamma$ and CO 2-0 lines, their flux ratios can be directly compared as they are not affected by extinction effects. The first ionization potential of sodium is low (5.1 eV), therefore it can be present near low-mass late-type stars, such as CTTS and EXors, and in regions which are shielded from ionizing photons around earlier-type stars, such as in disks. As hydrogen is neutral in the Na$^+$ region, a low value of the Br$\gamma$/Na\,{\sc{i}} ratio suggests low ionization and the presence of higher density regions. 
In the right panel of Fig. \ref{fig:ew_comparison_co_brgamma}, we plot the CO 2-0/Na\,{\sc{i}} ratio versus the Br$\gamma$/Na\,{\sc{i}} ratio for the same sample as used to compare the Br$\gamma$ and CO 2-0 equivalent widths. 
The CO 2-0/Na\,{\sc{i}} is in the range between 2 and 5 for most EXors, including Gaia23bab. Most Br$\gamma$/Na\,{\sc{i}} ratios are also low, which was interpreted by \citet{Lorenzetti2009} as it is because the role of the circumstellar disc dominates. 
Gaia23bab shows one of the highest CO 2-0/Na\,{\sc{i}} as well as Br$\gamma$/Na\,{\sc{i}} ratio among the plotted sample.
It is interesting to note that V899 Mon, a peculiar eruptive YSO with an emission line spectrum has a CO 2-0/Na\,{\sc{i}} ratio well above the range where most EXors are. In the case of the Class I type YSO IRAS 04239+2436, both the CO 2-0/Na\,{\sc{i}} and the Br$\gamma$/Na\,{\sc{i}} ratios are outside of the range that most EXors in the figure represent.

In addition to its spectral features and accretion rate, the recurrence of the outbursts also points to an EXor origin. 
In addition to the two well-defined outbursts seen in the light curves in Fig. \ref{fig:lightcurve}, another earlier outburst was identified in Pan-STARRS data between 2012 April and 2014 June \citep{Giannini2024}. As mentioned in Sect. \ref{sec:light_color} and in \citet{Giannini2024}, the two outbursts seen in the \textit{Gaia} light curve have a similar amplitude and color variation. A direct comparison of these two outbursts with the earlier seen in the Pan-STARRS data is not possible due to the low number of data points which do not allow to precisely derive the amplitude and timescale of that earlier outburst. However, based on the parameters given in Table 1 of \citet{Giannini2024}, this earlier outburst might also be comparable to the two outbursts seen in the \textit{Gaia} light curve. These two outbursts have different durations: comparing the FWHM values fitted to the light curves in Sect. \ref{sec:light_color}, the 2023 brightening was a factor of 1.5--2 longer than the one in 2017.

According to the recent definition of accretion-related events by \citet{Fischer2023}, the accretion events with an amplitude of 1--2.5 mag are referred to as bursts, while the accretion events with an amplitude above 2.5 mag are called outbursts. According to this definition, the brightening events of Gaia23bab are closer to bursts, as their amplitudes are only above 2.5 in ZTF bands (see Sect. \ref{sec:light_color} and \citealp{Giannini2024}). \citet{CruzSaenzdeMiera2023} present in their Fig. 1 the long-term light curve of EX Lupi, the prototype of EXors. During the $\sim$130 years covered by the light curve, only three outbursts (according to the \citet{Fischer2023} definition) are seen, however, there are many -- at least 19 -- bursts with similar amplitudes to those produced by Gaia23bab in 2017 and 2023. Other EXors with long-term light curves include VY Tau, which had 14 outbursts between 1940 and 1970 \citep{Herbig1990}, as well as V1118 Ori, which had at least five outbursts over 30 years \citep{Giannini2020}.
In order to better constrain the timescales, amplitude, and recurrence timescales of EXor (out)bursts, their long-term monitoring with ongoing and future surveys such as ZTF and the Legacy Survey of Space and Time (LSST) is important.

\section{Summary}
\label{sect_summary}

We analysed the physical properties of an EXor, Gaia23bab, which was recently discovered based on \textit{Gaia} Photometric Science Alerts. We studied its 2023 outburst using NIR spectra collected at three different epochs during the outburst. We also measured NIR and optical photometry during the outburst, and collected archival photometry, which included an earlier (2017) outburst of the target. Our main results can be summarized as follows:

ATLAS $o$ and $c$ data reveal a similar color evolution of the source during both the 2017 and the 2023 outbursts, which may partly be related to changes in circumstellar extinction. 

The spectra obtained at the three epochs during the 2023 outburst all show typical EXor signatures, with some apparent changes between the epochs, such as the non-detection of the CO bandhead and the only tentative detection of the Na\,{\sc{i}} doublet lines at the epoch during the fading of the source.

At two different epochs during the outburst we found the accretion rate to be consistent with $\sim$2$\times$10$^{-7}$ $M_\odot$ yr$^{-1}$. A similar value was found by \citet{Giannini2024} for a third epoch during the outburst, suggesting that the accretion rate did not change significantly during the outburst.

A comparison of the fluxes of the Brackett series to predictions of Case B theory resulted in a best fit temperature of 5000~K and a best fit electron density was $10^9$ cm$^{-3}$ at the NOT and LBT epochs, and temperatures of 7500~K and 10000~K and an electron density of $10^{10}$ cm$^{-3}$ at the IRTF epoch. These values are consistent with earlier works on EXors and other types of eruptive YSOs.

In comparison to the predictions of the \citet{KwanFischer2011} model, the observed Balmer decrement is consistent with temperatures of 5000 -- 12500~K and a hydrogen density of $10^8$ cm$^{-3}$.
At the LBT and IRTF epochs, the fluxes of the Paschen series are closest to a model with n$_H$ of $10^{11}$ cm$^{-3}$ and are consistent with temperatures of 10000 -- 12500~K.

The Ca\,{\sc{ii}} 8498 / O\,{\sc{i}} 8446 as well as the Ca\,{\sc{ii}} 8498 / Pa$\gamma$ line ratios are likely to trace a component with H densities of $\sim$10$^{12}$ cm$^{-3}$ and temperatures $\lesssim$7500 K, which may be related to the disk boundary layer.

Gaia23bab shows physical properties of a prototypical EXor, not only based on its accretion rate, but also based on the excitation conditions traced by its hydrogen lines. Monitoring its future outbursts will provide more information on how the physical conditions in a classical EXor change between outbursts.

\begin{acknowledgments}
We acknowledge the Hungarian National Research, Development and Innovation Office grant OTKA FK 146023.
We acknowledge support from the ESA PRODEX contract nr. 4000132054.
G. M. and Zs. N. were supported by the J\'anos Bolyai Research Scholarship of the Hungarian Academy of Sciences.

G.M is funded by the European Union’s Horizon 2020 research and innovation programme under grant agreement No. 101004141.

This work was also supported by the NKFIH NKKP grant ADVANCED 149943 and the NKFIH excellence grant TKP2021-NKTA-64. Project no.149943 has been implemented with the support provided by the Ministry of Culture and Innovation of Hungary from the National Research, Development and Innovation Fund, financed under the NKKP ADVANCED funding scheme.

Based on observations made with the Nordic Optical Telescope, owned in collaboration by the University of Turku and Aarhus University, and operated jointly by Aarhus University, the University of Turku and the University of Oslo, representing Denmark, Finland and Norway, the University of Iceland and Stockholm University at the Observatorio del Roque de los Muchachos, La Palma, Spain, of the Instituto de Astrofisica de Canarias.

This work was supported by the OPTICON-RadioNet Pilot (ORP) of the European Union's Horizon 2020 research and innovation programme under grant agreement No 101004719.

This work has made use of data from the Asteroid Terrestrial-impact Last Alert System (ATLAS) project. The Asteroid Terrestrial-impact Last Alert System (ATLAS) project is primarily funded to search for near Earth asteroids through NASA grants NN12AR55G, 80NSSC18K0284, and 80NSSC18K1575; byproducts of the NEO search include images and catalogs from the survey area. This work was partially funded by Kepler/K2 grant J1944/80NSSC19K0112 and HST GO-15889, and STFC grants ST/T000198/1 and ST/S006109/1. The ATLAS science products have been made possible through the contributions of the University of Hawaii Institute for Astronomy, the Queen’s University Belfast, the Space Telescope Science Institute, the South African Astronomical Observatory, and The Millennium Institute of Astrophysics (MAS), Chile.

We acknowledge ESA \textit{Gaia}, DPAC and the Photometric Science Alerts Team (\url{http://gsaweb.ast.cam.ac.uk/alerts}).

BHTOM.space has been based on the open-source TOM Toolkit by LCO and has been developed with funding from the European Union’s Horizon 2020 and Horizon Europe research and innovation programmes under grant agreements No 101004719 (OPTICON-RadioNet Pilot, ORP) and No 101131928 (ACME).

The operation of the RC80 telescope was supported by the project GINOP 2.3.2-15-2016-00033 of the Hungarian government, funded by the European Union.

This paper uses observations made at the Mount Suhora Astronomical Observatory, Poland.

This research was supported by the `SeismoLab' KKP-137523 \'Elvonal grant of the Hungarian Research, Development and Innovation Office (NKFIH).

Zs. B. was supported by the \'UNKP 22-2 New National Excellence Program of the Ministry for Culture and Innovation from the source of the National Research, Development and Innovation Fund.

B. S. and L. K. acknowledge the support of the Hungarian National Research, Development and Innovation Office grants KKP-143986 (\'Elvonal).
L. K. acknowledges the support of the Hungarian National Research, Development and Innovation Office grant OTKA PD-134784.

N. O. Sz. thanks the financial support provided by the undergraduate research assistant program of Konkoly Observatory.

Zs. M. Sz. acknowledges funding from a St. Leonards scholarship from the University of St. Andrews.
Zs. M. Sz. is a member of the International Max Planck Research School (IMPRS) for Astronomy and Astrophysics at the Universities of Bonn and Cologne.

F. C. S. M. received financial support from the European Research Council (ERC) under the European Union’s Horizon 2020 research and innovation programme (ERC Starting Grant "Chemtrip", grant agreement No 949278).

E. F. has been partially supported by project AYA2018-RTI-096188-B-I00 from the Spanish Agencia Estatal de Investigación and by Grant Agreement 101004719 of the EU project ORP.

\end{acknowledgments}

%

\vspace{5mm}
\facilities{NOT, LBT, IRTF, BHTOM.space}





\appendix

\section{Photometry}
\label{sec:appendix_phot}

\begin{table*}[!htb]
\centering
\caption{RC80 photometry of Gaia23bab.}
\label{tab:phot1}
\begin{tabular}{cccc|cccc}
\hline \hline
JD& $V$& $r$& $i$& JD& $V$& $r$& $i$\\
\hline

2460013.63&  17.98$\pm$0.15& 17.15$\pm$0.05& 15.84$\pm$0.03&
2460194.37&  19.56$\pm$0.21& 18.71$\pm$0.07& 17.15$\pm$0.03\\

2460020.61&  17.88$\pm$0.08& 17.00$\pm$0.03& 15.75$\pm$0.02&
2460195.42&  19.96$\pm$0.35& 18.59$\pm$0.09& 17.03$\pm$0.05\\

2460021.61&  17.98$\pm$0.33& 17.37$\pm$0.10& 16.07$\pm$0.04&
2460196.35&  19.77$\pm$0.21& 18.70$\pm$0.07& 17.10$\pm$0.03\\

2460022.62&  18.25$\pm$0.09& 17.38$\pm$0.04& 16.04$\pm$0.02&
2460197.39&  19.76$\pm$0.31& 18.73$\pm$0.09& 17.18$\pm$0.05\\

2460025.62&  18.20$\pm$0.10& 17.16$\pm$0.03& 15.81$\pm$0.02&
2460198.41&  19.79$\pm$0.30& 18.77$\pm$0.12& 17.16$\pm$0.04\\

2460026.58&  17.79$\pm$0.07& 16.87$\pm$0.02& 15.58$\pm$0.02&
2460200.38&  19.56$\pm$0.23& 18.85$\pm$0.11& 17.14$\pm$0.04\\

2460027.59&  18.25$\pm$1.35& 17.02$\pm$0.22& 15.64$\pm$0.09&
2460205.36&  20.05$\pm$0.84& 18.99$\pm$0.17& 17.18$\pm$0.05\\

2460029.58&  18.32$\pm$0.19& 17.29$\pm$0.05& 16.01$\pm$0.03&
2460217.36&  20.32$\pm$1.20& 18.88$\pm$0.19& 17.16$\pm$0.06\\

2460031.60&  18.30$\pm$0.13& 17.34$\pm$0.05& 15.99$\pm$0.02&
2460220.35&  19.87$\pm$0.45& 18.72$\pm$0.11& 17.16$\pm$0.04\\

2460032.61&  18.21$\pm$0.13& 17.31$\pm$0.04& 15.96$\pm$0.02&
2460224.34&  20.55$\pm$0.61& 19.34$\pm$0.15& 17.46$\pm$0.05\\

2460045.54&  18.09$\pm$0.14& 17.07$\pm$0.06& 15.78$\pm$0.03&
2460226.30&  19.41$\pm$0.45& 18.80$\pm$0.23& 17.24$\pm$0.06\\

2460047.52&  17.54$\pm$0.24& 16.81$\pm$0.06& 15.55$\pm$0.02&
2460236.29&  19.84$\pm$1.29& 19.32$\pm$0.63& 17.63$\pm$0.17\\

2460054.53&  17.82$\pm$0.08& 17.00$\pm$0.03& 15.67$\pm$0.05&
2460430.50&  20.97$\pm$0.45& 20.08$\pm$0.16& 17.75$\pm$0.04\\

2460055.53&  17.83$\pm$0.07& 16.93$\pm$0.03& 15.65$\pm$0.02&
2460431.53&  22.64$\pm$1.95& 21.39$\pm$0.47& 17.84$\pm$0.05\\

2460058.52&  18.06$\pm$0.11& 17.14$\pm$0.03& 15.83$\pm$0.02&
2460432.55&  21.87$\pm$1.06& 19.93$\pm$0.15& 17.80$\pm$0.09\\

2460060.49&  18.76$\pm$1.03& 17.14$\pm$0.12& 15.83$\pm$0.07&
2460449.47&  21.36$\pm$0.85& 22.16$\pm$1.38& 18.06$\pm$0.07\\

2460062.52&  17.85$\pm$0.08& 16.96$\pm$0.03& 15.70$\pm$0.02&
2460461.52&  21.71$\pm$0.78& 21.82$\pm$0.73& 17.95$\pm$0.09\\

2460064.56&  18.13$\pm$0.07& 17.20$\pm$0.04& 15.88$\pm$0.02&
2460478.50&  20.66$\pm$0.79& 20.01$\pm$0.24& 17.81$\pm$0.05\\

2460065.50&  18.06$\pm$0.09& 17.20$\pm$0.03& 15.85$\pm$0.02&
2460479.44&  20.80$\pm$0.38& 20.21$\pm$0.18& 17.81$\pm$0.04\\

2460069.50&  18.19$\pm$0.18& 17.19$\pm$0.07& 15.92$\pm$0.03&
2460480.40&  21.36$\pm$0.84& 22.31$\pm$1.84& 16.97$\pm$1.28\\

2460097.53&  18.03$\pm$0.05& 17.14$\pm$0.03& 15.80$\pm$0.02&
2460481.41&  20.49$\pm$0.46& 20.13$\pm$0.26& 17.68$\pm$0.05\\

2460098.50&  18.00$\pm$0.14& 17.14$\pm$0.05& 15.78$\pm$0.02&
2460486.49&  ...&            20.32$\pm$0.26& 17.82$\pm$0.06\\

2460099.51&  18.12$\pm$0.11& 17.32$\pm$0.04& 15.94$\pm$0.02&
2460497.49&  21.28$\pm$0.45& 20.21$\pm$0.15& 17.86$\pm$0.04\\

2460109.53&  18.14$\pm$0.05& 17.26$\pm$0.03& 15.96$\pm$0.02&
2460498.40&  21.26$\pm$0.46& 20.35$\pm$0.16& 17.82$\pm$0.03\\

2460110.48&  18.13$\pm$0.05& 17.25$\pm$0.03& 15.94$\pm$0.02&
2460500.46&  20.46$\pm$0.24& 19.73$\pm$0.11& 17.73$\pm$0.03\\

2460113.48&  18.58$\pm$0.08& 17.70$\pm$0.03& 16.31$\pm$0.02&
2460501.42&  21.22$\pm$0.51& 20.12$\pm$0.15& 17.87$\pm$0.04\\

2460114.55&  18.77$\pm$0.12& 17.68$\pm$0.06& 16.25$\pm$0.03&
2460502.44&  ...&            20.17$\pm$0.40& 17.80$\pm$0.09\\

2460115.47&  18.42$\pm$0.07& 17.49$\pm$0.03& 16.12$\pm$0.02&
2460503.45&  21.42$\pm$1.69&            ...& 17.72$\pm$0.26\\

2460116.51&  18.38$\pm$0.06& 17.45$\pm$0.03& 16.09$\pm$0.02&
2460505.44&  20.30$\pm$0.42& 19.36$\pm$0.24& 17.69$\pm$0.07\\

2460118.51&  17.68$\pm$0.05& 16.82$\pm$0.02& 15.55$\pm$0.01&
2460506.47&  20.13$\pm$0.23& 19.92$\pm$0.15& 17.67$\pm$0.05\\

2460122.50&  17.78$\pm$0.04& 16.89$\pm$0.02& 15.58$\pm$0.01&
2460507.53&  19.40$\pm$0.45& 20.69$\pm$0.62& 17.74$\pm$0.08\\

2460126.45&  17.95$\pm$0.10& 17.04$\pm$0.03& 15.72$\pm$0.03&
2460510.46&  20.05$\pm$0.36& 20.15$\pm$0.36& 17.75$\pm$0.07\\

2460128.39&  17.75$\pm$0.09& 16.88$\pm$0.04& 15.58$\pm$0.02&
2460511.49&  20.34$\pm$1.00& 19.45$\pm$0.28& 17.75$\pm$0.09\\

2460135.42&  18.83$\pm$0.10& 17.90$\pm$0.04& 16.55$\pm$0.02&
2460513.47&  18.98$\pm$0.38& 18.41$\pm$0.26& 17.37$\pm$0.20\\

2460137.56&  18.75$\pm$0.23& 17.67$\pm$0.09& 16.36$\pm$0.09&
2460514.53&  ...&            19.58$\pm$0.38& 17.80$\pm$0.11\\

2460141.40&  19.16$\pm$0.13& 18.18$\pm$0.05& 16.70$\pm$0.03&
2460516.47&             ...& 21.66$\pm$1.27& 17.93$\pm$0.11\\

2460143.41&  18.53$\pm$0.09& 17.63$\pm$0.03& 16.27$\pm$0.02&
2460518.49&  20.73$\pm$0.36& 20.87$\pm$0.32& 18.03$\pm$0.06\\

2460145.43&  18.75$\pm$0.10& 17.80$\pm$0.05& 16.40$\pm$0.02&
2460519.48&  21.66$\pm$0.77& 19.83$\pm$0.13& 17.94$\pm$0.07\\

2460146.38&  18.54$\pm$0.10& 17.75$\pm$0.04& 16.39$\pm$0.02&
2460520.49&  ...&            20.37$\pm$0.27& 17.97$\pm$0.07\\

2460171.42&  18.74$\pm$0.07& 17.80$\pm$0.03& 16.45$\pm$0.02&
2460521.52&  20.66$\pm$0.43& 20.19$\pm$0.21& 17.82$\pm$0.04\\

2460177.38&  18.92$\pm$0.12& 18.11$\pm$0.04& 16.68$\pm$0.02&
2460522.48&  21.45$\pm$0.66& 20.08$\pm$0.14& 17.90$\pm$0.05\\

2460178.38&  19.10$\pm$0.15& 18.19$\pm$0.06& 16.70$\pm$0.04&
2460523.52&  19.77$\pm$0.57& 19.80$\pm$0.53& 18.10$\pm$0.21\\

2460179.46&  19.53$\pm$0.57& 18.28$\pm$0.15& 16.75$\pm$0.05&
2460526.47&  20.80$\pm$0.71& 19.92$\pm$0.25& 18.11$\pm$0.08\\

2460180.39&  18.90$\pm$0.13& 17.97$\pm$0.05& 16.60$\pm$0.02&
2460528.51&  19.50$\pm$0.16& 20.02$\pm$0.18& 17.98$\pm$0.06\\

2460181.40&  19.14$\pm$0.12& 18.22$\pm$0.06& 16.80$\pm$0.04&
2460530.47&  19.09$\pm$1.72& 18.97$\pm$1.36& 17.06$\pm$0.58\\

2460182.37&  19.06$\pm$0.12& 18.18$\pm$0.05& 16.76$\pm$0.02&
2460532.51&  20.06$\pm$0.23& 20.82$\pm$0.32& 17.95$\pm$0.07\\

2460184.39&  19.06$\pm$0.23& 18.17$\pm$0.12& 16.81$\pm$0.07&
2460533.43&  21.19$\pm$0.44& 20.06$\pm$0.13& 17.91$\pm$0.04\\

2460193.37&  19.99$\pm$0.30& 19.02$\pm$0.10& 17.27$\pm$0.04&
2460534.42&             ...& 20.16$\pm$0.29& 17.93$\pm$0.08\\

\hline

\end{tabular}

\end{table*}	

\begin{table*}[!htbp]
\centering
\caption{RC80 photometry of Gaia23bab.}
\label{tab:phot2}
\begin{tabular}{cccc}
\hline \hline
JD& $V$& $r$& $i$\\
\hline
2460536.47&  22.11$\pm$1.41& 20.01$\pm$0.18&  17.88$\pm$0.09\\
2460537.45&             ...& 19.75$\pm$0.31&  17.73$\pm$0.07\\
2460538.45&  22.61$\pm$2.07& 20.17$\pm$0.18&  17.78$\pm$0.11\\
2460539.46&             ...& 20.02$\pm$0.20&  17.90$\pm$0.07\\
2460545.42&  21.35$\pm$0.99& 19.45$\pm$0.16&  17.71$\pm$0.08\\
2460546.41&  20.70$\pm$0.43& 19.83$\pm$0.15&  17.77$\pm$0.04\\
2460547.45&  20.26$\pm$0.44& 19.94$\pm$0.30&  17.89$\pm$0.07\\
2460548.44&  21.50$\pm$1.26& 21.86$\pm$1.68&  18.00$\pm$0.08\\
2460551.44&  20.40$\pm$0.53& 20.06$\pm$0.27&  18.98$\pm$0.14\\
2460552.44&  20.47$\pm$0.31& 20.01$\pm$0.17&  17.82$\pm$0.04\\
2460553.45&  21.73$\pm$1.21& 20.01$\pm$0.18&  18.04$\pm$0.09\\
2460554.43&  19.87$\pm$0.69& 20.25$\pm$0.52&  17.80$\pm$0.09\\
2460558.42&  21.50$\pm$1.11& 20.24$\pm$0.27&  17.88$\pm$0.06\\
\hline
\end{tabular}
\end{table*}	

\begin{table*}[h!]
\centering
\caption{Mt Suhora photometry of Gaia23bab.}
\label{tab:phot3}
\begin{tabular}{cccc}
\hline \hline
MJD& $V$& $R$& $I$\\
\hline
60020.15& 17.73$\pm$0.03& 16.41$\pm$0.02& 14.91$\pm$0.02\\
60021.13& 18.18$\pm$0.06& 16.97$\pm$0.06& 15.34$\pm$0.05\\
60062.07& 17.69$\pm$0.04& 16.52$\pm$0.05& 15.08$\pm$0.04\\
60076.05& 17.92$\pm$0.04& 16.67$\pm$0.04& 15.26$\pm$0.03\\ 
60092.03& 17.91$\pm$0.05& 16.50$\pm$0.02& 15.21$\pm$0.04\\
60139.97& 18.53$\pm$0.13& 17.32$\pm$0.08& 15.65$\pm$0.05\\
60140.03& 18.72$\pm$0.09& 17.49$\pm$0.07& 15.76$\pm$0.05\\
60144.01& 18.30$\pm$0.06& 16.98$\pm$0.05& 15.55$\pm$0.05\\
60145.00& 18.42$\pm$0.06& 17.12$\pm$0.05& 15.65$\pm$0.05\\
60145.98& 18.35$\pm$0.05& 17.08$\pm$0.03& 15.60$\pm$0.03\\
60153.88& 18.22$\pm$0.08& 16.79$\pm$0.03& 15.35$\pm$0.03\\
60164.85& 18.05$\pm$0.04& 16.58$\pm$0.02& 15.32$\pm$0.04\\
60166.89& 18.08$\pm$0.03& 16.81$\pm$0.03& 15.45$\pm$0.04\\
60170.88& 18.58$\pm$0.07& 17.38$\pm$0.08& 15.81$\pm$0.06\\
60171.83& 18.58$\pm$0.07& 17.18$\pm$0.06& 15.61$\pm$0.03\\
60179.97& 18.61$\pm$0.11& 17.26$\pm$0.05& 15.52$\pm$0.04\\
60180.97& 18.96$\pm$0.08& 17.44$\pm$0.03& 15.67$\pm$0.03\\
60193.83& 19.56$\pm$0.09& 18.00$\pm$0.05& 16.17$\pm$0.05\\
60215.86& ...& 17.93$\pm$0.10& 15.98$\pm$0.04\\
60216.78& ...& 18.02$\pm$0.08& 15.95$\pm$0.04\\
60221.77& 19.79$\pm$0.27& 18.02$\pm$0.07& 16.23$\pm$0.04\\
60234.75& 19.90$\pm$0.15& 18.32$\pm$0.08& 16.59$\pm$0.07\\
60255.72& 20.06$\pm$0.19& 18.23$\pm$0.07& 16.13$\pm$0.06\\
60497.92& 21.18$\pm$0.29& 19.07$\pm$0.11& 16.79$\pm$0.10\\
60528.86& 20.71$\pm$0.15& 19.25$\pm$0.10& 17.10$\pm$0.12\\ 
\hline
\end{tabular}
\end{table*}	

\section{Line identification}
\label{line_ident}

\begin{figure*}[h]
\centering
\includegraphics[width=14cm]{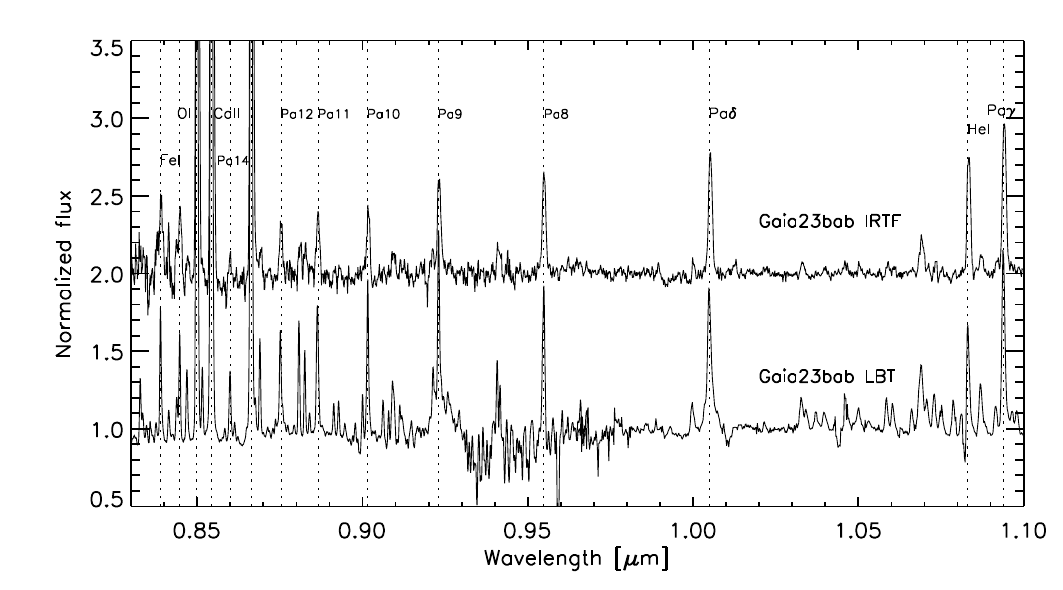}
\includegraphics[width=14cm]{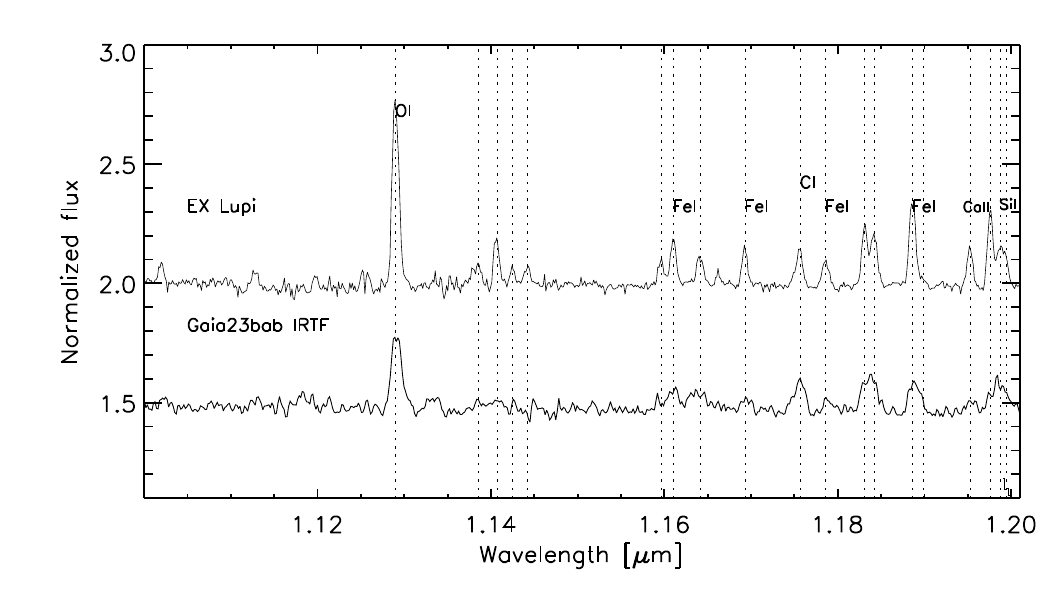}
\includegraphics[width=14cm]{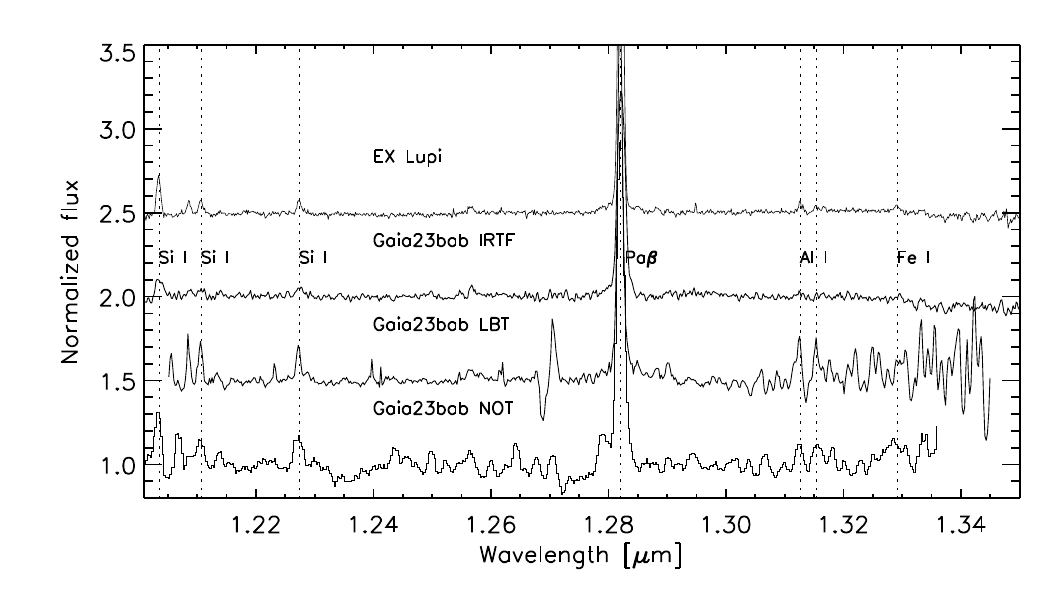}
\caption{Comparison of the NOT and IRTF spectra to those of EX Lupi during its outburst \citep{Kospal2011}.}
\label{fig:line_ident_1}
\end{figure*}

\begin{figure*}[h]
\centering
\includegraphics[width=14cm]{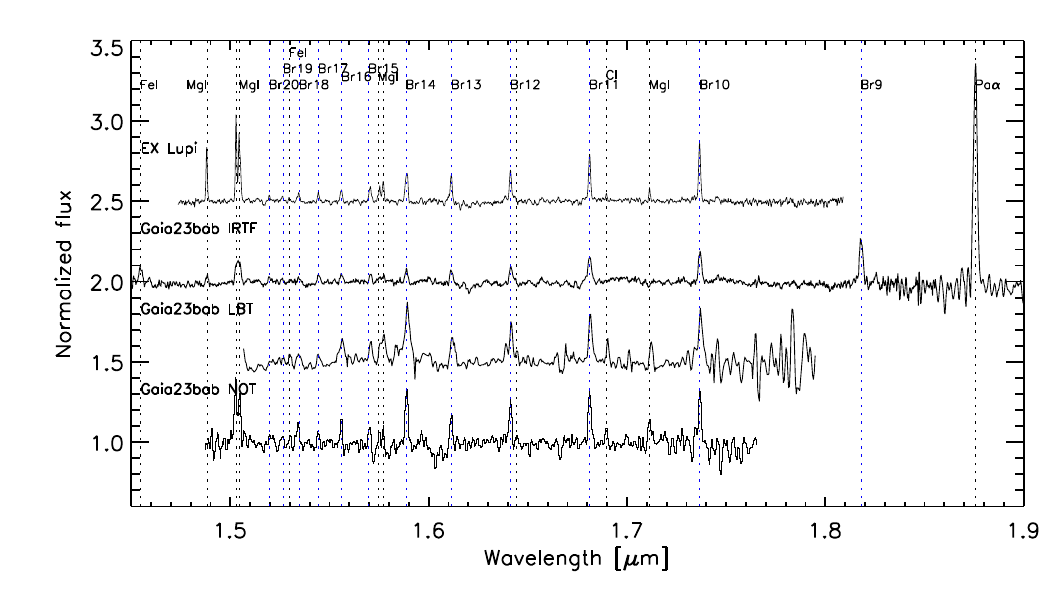}
\includegraphics[width=14cm]{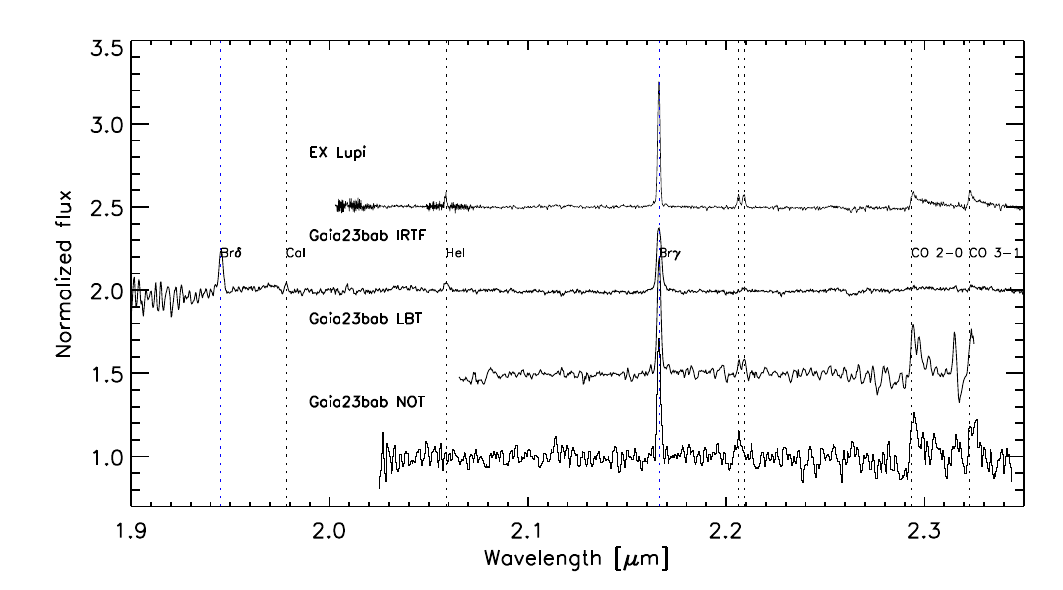}
\caption{Comparison of the NOT and IRTF spectra to those of EX Lupi during its outburst \citep{Kospal2011}.}
\label{fig:line_ident_2}
\end{figure*}

\begin{figure*}[h]
\centering
\includegraphics[width=7cm]{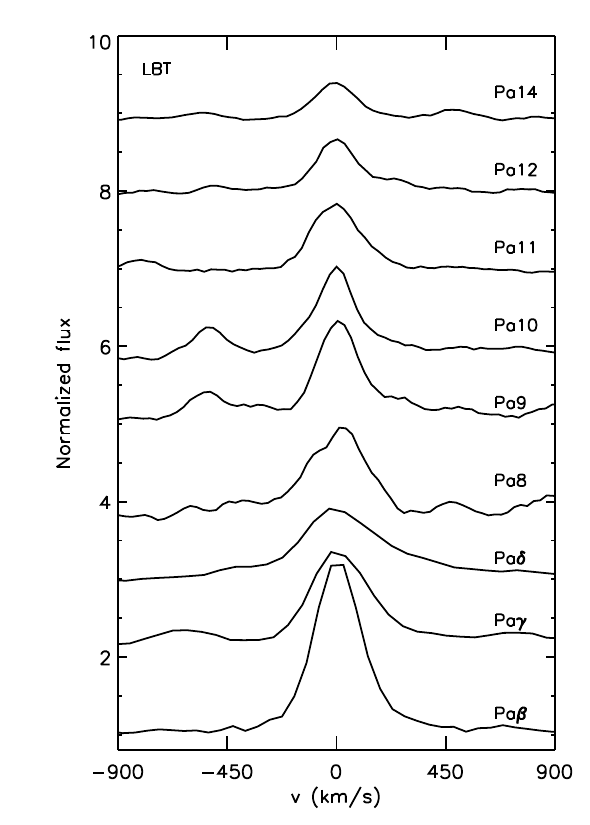}
\includegraphics[width=7cm]{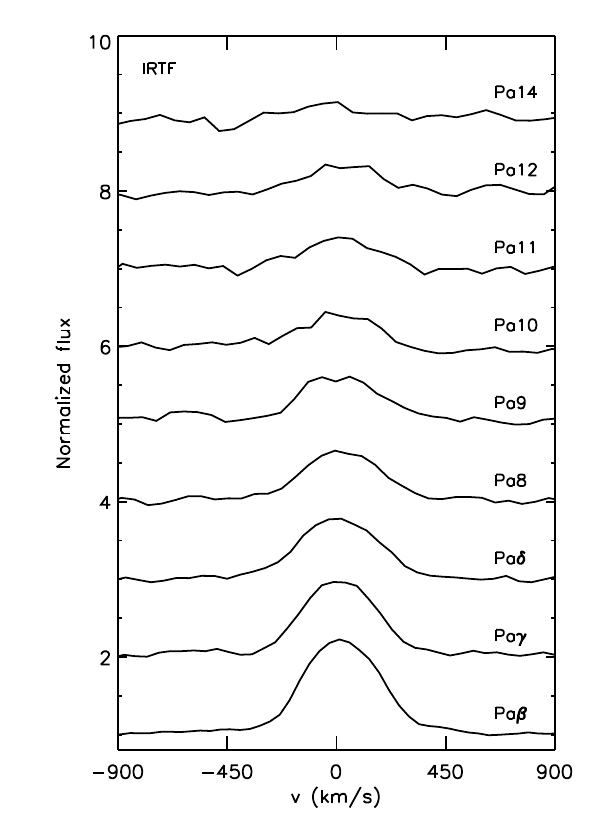}
\caption{The H\,{\sc{i}} Paschen lines detected in the spectra of Gaia23bab.}
\label{fig:pa_series}
\end{figure*}

\begin{figure*}[h]
\includegraphics[width=6cm]{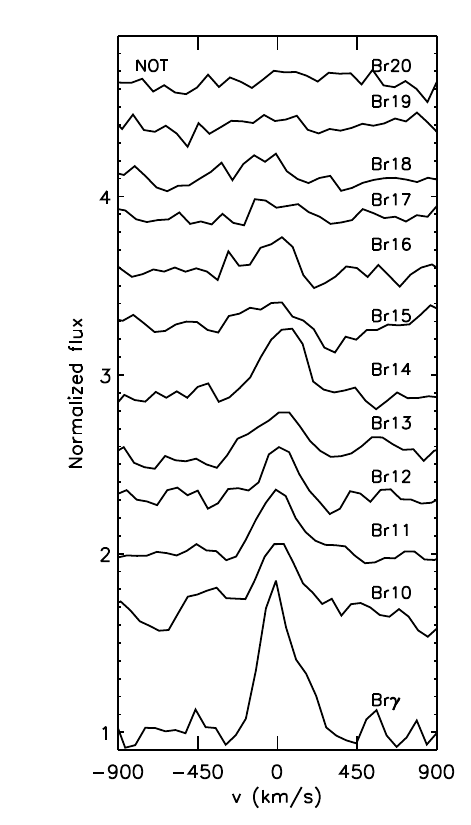}
\includegraphics[width=6cm]{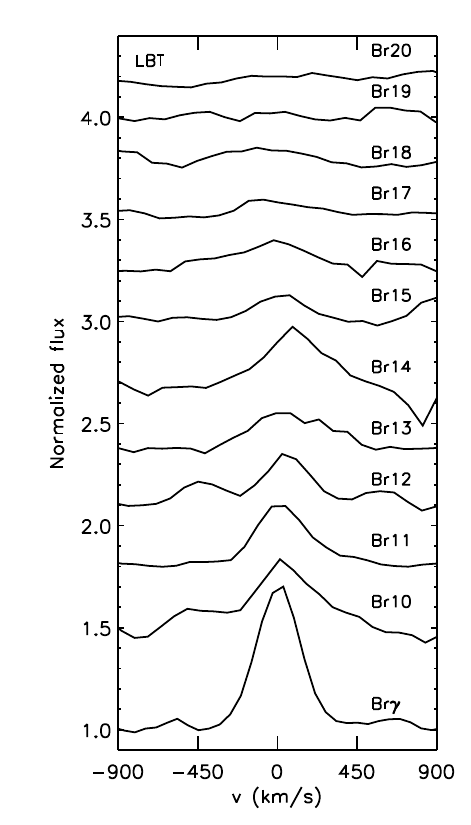}
\includegraphics[width=6cm]{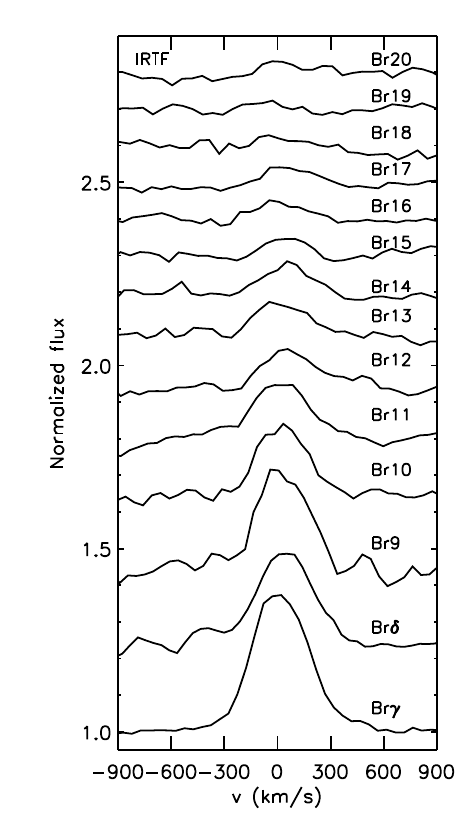}
\caption{The H\,{\sc{i}} Brackett lines detected in the spectra of Gaia23bab.}
\label{fig:br_series}
\end{figure*}

\section{Line parameters}

\begin{table}[h]
\centering
\caption{Equivalent widths of the spectral features detected at the LBT and IRTF epochs. The lines marked with $*$ are most likely blended at least at one of the epochs. Lines for which no equivalent widths are provided are either not covered by the data (--).}
\label{tab:spec_lines1}
\begin{tabular}{lcrr}
\hline \hline
Species& $\lambda_{\rm{tab}}$ ($\mu$m)& EW(LBT) (\AA)& EW(IRTF) (\AA)\\
\hline
Fe I&             0.8388&           $-4.8\pm0.2$&    $-7.5\pm0.5$\\
O I&              0.8446&           $-3.5\pm0.1$&    $-5.0\pm0.5$\\
Ca II&            0.8498&           $-50.0\pm1.0$&   $-62.0\pm2.0$\\
Ca II&            0.8542&           $-54.0\pm1.0$&   $-69.0\pm1.0$\\
Ca II&            0.8662&           $-53.5\pm2.0$&   $-62.0\pm2.0$\\
H I (Pa14)&       0.8601&           $-0.9\pm0.3$&    $-0.9\pm0.1$\\
H I (Pa13)*&      0.8667&           $-1.9\pm0.4$&    $-2.0\pm0.3$\\
H I (Pa12)&       0.8753&           $-4.2\pm0.3$&    $-4.1\pm0.3$\\
H I (Pa11)&       0.8865&           $-6.1\pm0.3$&    $-4.7\pm0.5$\\
H I (Pa10)&       0.9015&           $-6.2\pm0.3$&    $-4.9\pm0.3$\\
H I (Pa9)*&       0.9232&           $-9.4\pm0.5$&    $-7.5\pm0.1$\\
H I (Pa8)&        0.9546&           $-8.8\pm0.2$&    $-7.5\pm0.3$\\
H I (Pa$\delta$)& 1.0052&           $-12.5\pm0.6$&   $-11.3\pm0.5$\\
He I&             1.0830&           $-7.0\pm0.4$&    $-10.0\pm0.1$\\
H I (Pa$\gamma$)& 1.0941&           $-13.6\pm0.5$&   $-13.5\pm0.3$\\
O I&              1.1290&           --&               $-4.7\pm0.3$\\
Fe I&             1.1610&           --&               $-1.1\pm0.1$\\
Fe I&             1.1641&           --&               $-1.9\pm0.1$\\
Fe I&             1.1693&           --&               $-1.0\pm0.2$\\
C I&              1.1757&           --&               $-1.9\pm0.1$\\
Fe I&             1.1786&           --&               $-0.8\pm0.2$\\
Mg I&             1.1831&           --&               $-3.1\pm0.1$\\
\hline
\end{tabular}
\end{table}

\begin{table}[h]
\centering
\caption{Equivalent widths of the spectral features detected at the NOT, LBT, and IRTF epochs. The lines marked with $*$ are most likely blended at least at one of the epochs. Lines for which no equivalent widths are provided are either not covered by the data (--) or cannot be derived as it is either not detected or is blended (...).}
\label{tab:spec_lines2}
\begin{tabular}{lcrrr}
\hline \hline
Species& $\lambda_{\rm{tab}}$ ($\mu$m)& EW(NOT) (\AA)& EW(LBT) (\AA)& EW(IRTF) (\AA)\\
\hline
Si I&             1.2035&  $-4.4\pm0.1$&   $-2.2\pm0.1$&     $-1.8\pm0.2$\\
Si I&             1.2274&  $-3.0\pm0.2$&   $-1.9\pm0.2$&     $-0.9\pm0.1$\\
H I (Pa$\beta$)&  1.2821& $-31.0\pm2.0$&   $-25.0\pm0.6$&   $-21.5\pm0.2$\\
Al I&             1.3127&  $-2.9\pm0.2$&   $-2.5\pm0.2$&    ...\\
Al I&             1.3154&  $-3.1\pm0.2$&   $-2.4\pm0.2$&    ...\\
Fe I&             1.4547&            --&   --&              $-1.9\pm0.2$\\
Mg I&             1.4882&            --&   --&              $-0.9\pm0.1$\\
Mg I&             1.5029&  $-5.4\pm0.3$&   --&              ...\\
Mg I&             1.5044&  $-5.6\pm0.3$&   --&              ...\\
H I (Br20)&       1.5195&  $-1.2\pm0.2$&   $-0.8\pm0.2$&    $-0.4\pm0.1$\\
H I (Br19)&       1.5265&  $-1.3\pm0.2$&   $-1.3\pm0.2$&    $-0.5\pm0.1$\\
Fe I&             1.5299&  $-1.6\pm0.1$&   $-0.6\pm0.2$&    $-0.5\pm0.1$\\
H I (Br18)&       1.5346&  $-1.7\pm0.2$&   $-1.8\pm0.2$&    $-0.6\pm0.1$\\
H I (Br17)&       1.5443&  $-1.8\pm0.2$&   $-1.9\pm0.2$&    $-0.9\pm0.1$\\
H I (Br16)&       1.5561&  $-2.3\pm0.2$&   $-2.8\pm0.2$&    $-1.0\pm0.1$\\
H I (Br15)&       1.5705&  $-2.6\pm0.2$&   $-2.8\pm0.3$&    $-1.1\pm0.1$\\
Mg I*&            1.5745&  $-2.1\pm0.2$&            ...&    ...\\
Mg I*&            1.5770&  $-1.8\pm0.2$&            ...&    ...\\
H I (Br14)*&      1.5885&  $-6.3\pm0.2$&  $-10.3\pm0.3$&    $-1.8\pm0.1$\\  
H I (Br13)&       1.6114&  $-4.0\pm0.2$&   $-5.1\pm0.2$&    $-1.8\pm0.1$\\
H I (Br12)*&      1.6412&  $-4.6\pm0.1$&   $-5.5\pm0.3$&    $-2.4\pm0.2$\\
H I (Br11)&       1.6811&  $-5.2\pm0.2$&   $-6.3\pm0.2$&    $-3.2\pm0.2$\\
H I (Br10)&       1.7367&  $-6.4\pm0.2$&   $-7.6\pm0.3$&    $-4.3\pm0.2$\\
H I (Br9)&        1.8179&            --&             --&    $-6.7\pm0.2$\\
H I (Pa$\alpha$)& 1.8756&            --&             --&    $-34.0\pm1.0$\\
H I (Br$\delta$)& 1.9451&            --&             --&     $-6.8\pm0.2$\\
Ca I&             1.9782&            --&             --&     $-0.9\pm0.1$\\
He I&             2.0587&           ...&             --&     $-1.2\pm0.1$\\
Na I&             2.2062&           ...&   $-1.6\pm0.2$&              ...\\
Na I&             2.2090&           ...&   $-2.2\pm0.2$&              ...\\
H I (Br$\gamma$)& 2.1661& $-14.1\pm0.5$&  $-16.5\pm0.4$&    $-11.0\pm1.0$\\
CO 2-0&           2.2992& $-15.0\pm0.5$&  $-15.2\pm0.5$&              ...\\
\hline
\end{tabular}
\end{table}


\bibliography{gaia23bab}{}

\begin{thebibliography}{}
\expandafter\ifx\csname natexlab\endcsname\relax\def\natexlab#1{#1}\fi
\providecommand{\url}[1]{\href{#1}{#1}}
\providecommand{\dodoi}[1]{doi:~\href{http://doi.org/#1}{\nolinkurl{#1}}}
\providecommand{\doeprint}[1]{\href{http://ascl.net/#1}{\nolinkurl{http://ascl.net/#1}}}
\providecommand{\doarXiv}[1]{\href{https://arxiv.org/abs/#1}{\nolinkurl{https://arxiv.org/abs/#1}}}

\bibitem[{{Alcal{\'a}} {et~al.}(2017){Alcal{\'a}}, {Manara}, {Natta}, {Frasca},
  {Testi}, {Nisini}, {Stelzer}, {Williams}, {Antoniucci}, {Biazzo}, {Covino},
  {Esposito}, {Getman}, \& {Rigliaco}}]{Alcala2017}
{Alcal{\'a}}, J.~M., {Manara}, C.~F., {Natta}, A., {et~al.} 2017, \aap, 600,
  A20, \dodoi{10.1051/0004-6361/201629929}

\bibitem[{{Audard} {et~al.}(2014){Audard}, {{\'A}brah{\'a}m}, {Dunham},
  {Green}, {Grosso}, {Hamaguchi}, {Kastner}, {K{\'o}sp{\'a}l}, {Lodato},
  {Romanova}, {Skinner}, {Vorobyov}, \& {Zhu}}]{Audard2014}
{Audard}, M., {{\'A}brah{\'a}m}, P., {Dunham}, M.~M., {et~al.} 2014, in
  Protostars and Planets VI, ed. H.~{Beuther}, R.~S. {Klessen}, C.~P.
  {Dullemond}, \& T.~{Henning}, 387,
  \dodoi{10.2458/azu\_uapress\_9780816531240-ch017}

\bibitem[{{Azevedo} {et~al.}(2006){Azevedo}, {Calvet}, {Hartmann}, {Folha},
  {Gameiro}, \& {Muzerolle}}]{Azevedo2006}
{Azevedo}, R., {Calvet}, N., {Hartmann}, L., {et~al.} 2006, \aap, 456, 225,
  \dodoi{10.1051/0004-6361:20054315}

\bibitem[{{Bary} {et~al.}(2008){Bary}, {Matt}, {Skrutskie}, {Wilson},
  {Peterson}, \& {Nelson}}]{Bary2008}
{Bary}, J.~S., {Matt}, S.~P., {Skrutskie}, M.~F., {et~al.} 2008, \apj, 687,
  376, \dodoi{10.1086/591487}

\bibitem[{{Bessell} \& {Brett}(1988)}]{BessellBrett1988}
{Bessell}, M.~S., \& {Brett}, J.~M. 1988, \pasp, 100, 1134,
  \dodoi{10.1086/132281}

\bibitem[{{Cardelli} {et~al.}(1989){Cardelli}, {Clayton}, \&
  {Mathis}}]{Cardelli1989}
{Cardelli}, J.~A., {Clayton}, G.~C., \& {Mathis}, J.~S. 1989, \apj, 345, 245,
  \dodoi{10.1086/167900}

\bibitem[{{Connelley} \& {Greene}(2010)}]{ConnelleyGreene2010}
{Connelley}, M.~S., \& {Greene}, T.~P. 2010, \aj, 140, 1214,
  \dodoi{10.1088/0004-6256/140/5/1214}

\bibitem[{{Cruz-S{\'a}enz de Miera} {et~al.}(2022){Cruz-S{\'a}enz de Miera},
  {K{\'o}sp{\'a}l}, {{\'A}brah{\'a}m}, {Park}, {Nagy}, {Siwak}, {Kun},
  {Fiorellino}, {Szab{\'o}}, {Antoniucci}, {Giannini}, {Nisini}, {Szabados},
  {Kriskovics}, {Ordasi}, {Szak{\'a}ts}, {Vida}, {Vink{\'o}}, {Zieli{\'n}ski},
  {Wyrzykowski}, {Garc{\'\i}a-{\'A}lvarez}, {Dr{\'o}{\.z}d{\.z}}, {Og{\l}oza},
  \& {Sonbas}}]{CruzSaenzdeMiera2022}
{Cruz-S{\'a}enz de Miera}, F., {K{\'o}sp{\'a}l}, {\'A}., {{\'A}brah{\'a}m}, P.,
  {et~al.} 2022, \apj, 927, 125, \dodoi{10.3847/1538-4357/ac477f}

\bibitem[{{Cruz-S{\'a}enz de Miera} {et~al.}(2023){Cruz-S{\'a}enz de Miera},
  {K{\'o}sp{\'a}l}, {Abrah{\'a}m}, {Claes}, {Manara}, {Wendeborn},
  {Fiorellino}, {Giannini}, {Nisini}, {Sicilia-Aguilar}, {Campbell-White},
  {Alcal{\'a}}, {Banzatti}, {Szab{\'o}}, {Lykou}, {Antoniucci}, {Varga},
  {Siwak}, {Park}, {Nagy}, \& {Kun}}]{CruzSaenzdeMiera2023}
{Cruz-S{\'a}enz de Miera}, F., {K{\'o}sp{\'a}l}, {\'A}., {Abrah{\'a}m}, P.,
  {et~al.} 2023, \aap, 678, A88, \dodoi{10.1051/0004-6361/202347063}

\bibitem[{{Cushing} {et~al.}(2004){Cushing}, {Vacca}, \&
  {Rayner}}]{Cushing2004}
{Cushing}, M.~C., {Vacca}, W.~D., \& {Rayner}, J.~T. 2004, \pasp, 116, 362,
  \dodoi{10.1086/382907}

\bibitem[{{Cutri} {et~al.}(2003){Cutri}, {Skrutskie}, {van Dyk}, {Beichman},
  {Carpenter}, {Chester}, {Cambresy}, {Evans}, {Fowler}, {Gizis}, {Howard},
  {Huchra}, {Jarrett}, {Kopan}, {Kirkpatrick}, {Light}, {Marsh}, {McCallon},
  {Schneider}, {Stiening}, {Sykes}, {Weinberg}, {Wheaton}, {Wheelock}, \&
  {Zacarias}}]{cutri2003}
{Cutri}, R.~M., {Skrutskie}, M.~F., {van Dyk}, S., {et~al.} 2003, {2MASS All
  Sky Catalog of point sources.} (NASA/IPAC Infrared Science Archive)

\bibitem[{{Edwards} {et~al.}(2013){Edwards}, {Kwan}, {Fischer}, {Hillenbrand},
  {Finn}, {Fedorenko}, \& {Feng}}]{Edwards2013}
{Edwards}, S., {Kwan}, J., {Fischer}, W., {et~al.} 2013, \apj, 778, 148,
  \dodoi{10.1088/0004-637X/778/2/148}

\bibitem[{{Fiorellino} {et~al.}(2024){Fiorellino}, {{\'A}brah{\'a}m},
  {K{\'o}sp{\'a}l}, {Kun}, {Alcal{\'a}}, {Caratti o Garatti}, {Cruz-S{\'a}enz
  de Miera}, {Garc{\'\i}a-{\'A}lvarez}, {Giannini}, {Park}, {Siwak},
  {Szil{\'a}gyi}, {Covino}, {Marton}, {Nagy}, {Nisini}, {Marianna Szab{\'o}},
  {Bora}, {Cseh}, {Kalup}, {Krezinger}, {Kriskovics}, {Og{\l}oza}, {P{\'a}l},
  {S{\'o}dor}, {Sonbas}, {Szak{\'a}ts}, {Vida}, {Vink{\'o}}, {Wyrzykowski}, \&
  {Zielinski}}]{Fiorellino2024}
{Fiorellino}, E., {{\'A}brah{\'a}m}, P., {K{\'o}sp{\'a}l}, {\'A}., {et~al.}
  2024, \aap, 686, A160, \dodoi{10.1051/0004-6361/202347777}

\bibitem[{{Fischer} {et~al.}(2023){Fischer}, {Hillenbrand}, {Herczeg},
  {Johnstone}, {Kospal}, \& {Dunham}}]{Fischer2023}
{Fischer}, W.~J., {Hillenbrand}, L.~A., {Herczeg}, G.~J., {et~al.} 2023, in
  Astronomical Society of the Pacific Conference Series, Vol. 534, Protostars
  and Planets VII, ed. S.~{Inutsuka}, Y.~{Aikawa}, T.~{Muto}, K.~{Tomida}, \&
  M.~{Tamura}, 355, \dodoi{10.48550/arXiv.2203.11257}

\bibitem[{{Gaia Collaboration} {et~al.}(2023){Gaia Collaboration}, {Vallenari},
  {Brown}, {Prusti}, {de Bruijne}, {Arenou}, {Babusiaux}, {Biermann},
  {Creevey}, {Ducourant}, \& et~al.}]{GaiaCollaboration2023}
{Gaia Collaboration}, {Vallenari}, A., {Brown}, A.~G.~A., {et~al.} 2023, \aap,
  674, A1, \dodoi{10.1051/0004-6361/202243940}

\bibitem[{{Ghosh} {et~al.}(2022){Ghosh}, {Sharma}, {Ninan}, {Ojha}, {Bhatt},
  {Kanodia}, {Mahadevan}, {Stefansson}, {Yadav}, {Gour}, {Pandey}, {Sinha},
  {Panwar}, {Wisniewski}, {Ca{\~n}as}, {Lin}, {Roy}, {Hearty}, {Ramsey},
  {Robertson}, \& {Schwab}}]{Ghosh2022}
{Ghosh}, A., {Sharma}, S., {Ninan}, J.~P., {et~al.} 2022, \apj, 926, 68,
  \dodoi{10.3847/1538-4357/ac41c2}

\bibitem[{{Giannini} {et~al.}(2020){Giannini}, {Giunta}, {Lorenzetti},
  {Altavilla}, {Antoniucci}, {Strafella}, \& {Testa}}]{Giannini2020}
{Giannini}, T., {Giunta}, A., {Lorenzetti}, D., {et~al.} 2020, \aap, 637, A83,
  \dodoi{10.1051/0004-6361/202037695}

\bibitem[{{Giannini} {et~al.}(2022){Giannini}, {Giunta}, {Gangi}, {Carini},
  {Lorenzetti}, {Antoniucci}, {Caratti o Garatti}, {Cassar{\'a}}, {Nisini},
  {Rossi}, {Testa}, \& {Vitali}}]{Giannini2022}
{Giannini}, T., {Giunta}, A., {Gangi}, M., {et~al.} 2022, \apj, 929, 129,
  \dodoi{10.3847/1538-4357/ac5a49}

\bibitem[{{Giannini} {et~al.}(2024){Giannini}, {Schisano}, {Nisini},
  {{\'A}brah{\'a}m}, {Antoniucci}, {Biazzo}, {Cruz-S{\'a}enz de Miera},
  {Fiorellino}, {Gangi}, {K{\'o}sp{\'a}l}, {Kuhn}, {Marini}, {Nagy}, \&
  {Paris}}]{Giannini2024}
{Giannini}, T., {Schisano}, E., {Nisini}, B., {et~al.} 2024, \apj, 967, 41,
  \dodoi{10.3847/1538-4357/ad39e2}

\bibitem[{{Greene} \& {Lada}(1996)}]{GreeneLada1996}
{Greene}, T.~P., \& {Lada}, C.~J. 1996, \apj, 461, 345, \dodoi{10.1086/177061}

\bibitem[{{Hartmann} {et~al.}(1998){Hartmann}, {Calvet}, {Gullbring}, \&
  {D'Alessio}}]{Hartmann1998}
{Hartmann}, L., {Calvet}, N., {Gullbring}, E., \& {D'Alessio}, P. 1998, \apj,
  495, 385, \dodoi{10.1086/305277}

\bibitem[{{Hartmann} \& {Kenyon}(1996)}]{HartmannKenyon1996}
{Hartmann}, L., \& {Kenyon}, S.~J. 1996, \araa, 34, 207,
  \dodoi{10.1146/annurev.astro.34.1.207}

\bibitem[{{Heinze} {et~al.}(2018){Heinze}, {Tonry}, {Denneau}, {Flewelling},
  {Stalder}, {Rest}, {Smith}, {Smartt}, \& {Weiland}}]{Heinze2018}
{Heinze}, A.~N., {Tonry}, J.~L., {Denneau}, L., {et~al.} 2018, \aj, 156, 241,
  \dodoi{10.3847/1538-3881/aae47f}

\bibitem[{{Henden} {et~al.}(2015){Henden}, {Levine}, {Terrell}, \&
  {Welch}}]{henden2015}
{Henden}, A.~A., {Levine}, S., {Terrell}, D., \& {Welch}, D.~L. 2015, in
  American Astronomical Society Meeting Abstracts, Vol. 225, American
  Astronomical Society Meeting Abstracts \#225, 336.16

\bibitem[{{Herbig}(1977)}]{Herbig1977}
{Herbig}, G.~H. 1977, \apj, 217, 693, \dodoi{10.1086/155615}

\bibitem[{{Herbig}(1990)}]{Herbig1990}
---. 1990, \apj, 360, 639, \dodoi{10.1086/169151}

\bibitem[{{Herbig}(2008)}]{Herbig2008}
---. 2008, \aj, 135, 637, \dodoi{10.1088/0004-6256/135/2/637}

\bibitem[{{Hillenbrand} {et~al.}(2019){Hillenbrand}, {Reipurth}, {Connelley},
  {Cutri}, \& {Isaacson}}]{Hillenbrand2019}
{Hillenbrand}, L.~A., {Reipurth}, B., {Connelley}, M., {Cutri}, R.~M., \&
  {Isaacson}, H. 2019, \aj, 158, 240, \dodoi{10.3847/1538-3881/ab4e16}

\bibitem[{{Hillenbrand} {et~al.}(2018){Hillenbrand}, {Contreras Pe{\~n}a},
  {Morrell}, {Naylor}, {Kuhn}, {Cutri}, {Rebull}, {Hodgkin}, {Froebrich}, \&
  {Mainzer}}]{Hillenbrand2018}
{Hillenbrand}, L.~A., {Contreras Pe{\~n}a}, C., {Morrell}, S., {et~al.} 2018,
  \apj, 869, 146, \dodoi{10.3847/1538-4357/aaf414}

\bibitem[{{Hodapp} {et~al.}(2019){Hodapp}, {Reipurth}, {Pettersson}, {Tonry},
  {Denneau}, {Vallely}, {Shappee}, {Armstrong}, {Connelley}, {Kochanek},
  {Fausnaugh}, {Chini}, {Haas}, \& {Sobrino Figaredo}}]{Hodapp2019}
{Hodapp}, K.~W., {Reipurth}, B., {Pettersson}, B., {et~al.} 2019, \aj, 158,
  241, \dodoi{10.3847/1538-3881/ab471a}

\bibitem[{{Hodapp} {et~al.}(2020){Hodapp}, {Denneau}, {Tucker}, {Shappee},
  {Huber}, {Payne}, {Do}, {Lin}, {Connelley}, {Varricatt}, {Tonry}, {Chambers},
  \& {Magnier}}]{Hodapp2020}
{Hodapp}, K.~W., {Denneau}, L., {Tucker}, M., {et~al.} 2020, \aj, 160, 164,
  \dodoi{10.3847/1538-3881/abad96}

\bibitem[{{Hodgkin} {et~al.}(2021){Hodgkin}, {Harrison}, {Breedt}, {Wevers},
  {Rixon}, {Delgado}, {Yoldas}, {Kostrzewa-Rutkowska}, {Wyrzykowski}, {van
  Leeuwen}, {Blagorodnova}, {Campbell}, {Eappachen}, {Fraser}, {Ihanec},
  {Koposov}, {Kruszy{\'n}ska}, {Marton}, {Rybicki}, {Brown}, {Burgess},
  {Busso}, {Cowell}, {De Angeli}, {Diener}, {Evans}, {Gilmore}, {Holland},
  {Jonker}, {van Leeuwen}, {Mignard}, {Osborne}, {Portell}, {Prusti},
  {Richards}, {Riello}, {Seabroke}, {Walton}, {{\'A}brah{\'a}m}, {Altavilla},
  {Baker}, {Bastian}, {O'Brien}, {de Bruijne}, {Butterley}, {Carrasco},
  {Casta{\~n}eda}, {Clark}, {Clementini}, {Copperwheat}, {Cropper},
  {Damljanovic}, {Davidson}, {Davis}, {Dennefeld}, {Dhillon}, {Dolding},
  {Dominik}, {Esquej}, {Eyer}, {Fabricius}, {Fridman}, {Froebrich}, {Garralda},
  {Gomboc}, {Gonz{\'a}lez-Vidal}, {Guerra}, {Hambly}, {Hardy}, {Holl},
  {Hourihane}, {Japelj}, {Kann}, {Kiss}, {Knigge}, {Kolb}, {Komossa},
  {K{\'o}sp{\'a}l}, {Kov{\'a}cs}, {Kun}, {Leto}, {Lewis}, {Littlefair},
  {Mahabal}, {Mundell}, {Nagy}, {Padeletti}, {Palaversa}, {Pigulski},
  {Pretorius}, {van Reeven}, {Ribeiro}, {Roelens}, {Rowell}, {Schartel},
  {Scholz}, {Schwope}, {Sip{\H{o}}cz}, {Smartt}, {Smith}, {Serraller},
  {Steeghs}, {Sullivan}, {Szabados}, {Szegedi-Elek}, {Tisserand}, {Tomasella},
  {van Velzen}, {Whitelock}, {Wilson}, \& {Young}}]{Hodgkin2021}
{Hodgkin}, S.~T., {Harrison}, D.~L., {Breedt}, E., {et~al.} 2021, \aap, 652,
  A76, \dodoi{10.1051/0004-6361/202140735}

\bibitem[{{Hummer} \& {Storey}(1987)}]{HummerStorey1987}
{Hummer}, D.~G., \& {Storey}, P.~J. 1987, \mnras, 224, 801,
  \dodoi{10.1093/mnras/224.3.801}

\bibitem[{{Jayasinghe} {et~al.}(2019){Jayasinghe}, {Stanek}, {Kochanek},
  {Shappee}, {Holoien}, {Thompson}, {Prieto}, {Dong}, {Pawlak}, {Pejcha},
  {Shields}, {Pojmanski}, {Otero}, {Hurst}, {Britt}, \&
  {Will}}]{jayasinghe2019}
{Jayasinghe}, T., {Stanek}, K.~Z., {Kochanek}, C.~S., {et~al.} 2019, \mnras,
  485, 961, \dodoi{10.1093/mnras/stz444}

\bibitem[{{K{\'o}sp{\'a}l} {et~al.}(2011){K{\'o}sp{\'a}l}, {{\'A}brah{\'a}m},
  {Goto}, {Reg{\'a}ly}, {Dullemond}, {Henning}, {Juh{\'a}sz},
  {Sicilia-Aguilar}, \& {van den Ancker}}]{Kospal2011}
{K{\'o}sp{\'a}l}, {\'A}., {{\'A}brah{\'a}m}, P., {Goto}, M., {et~al.} 2011,
  \apj, 736, 72, \dodoi{10.1088/0004-637X/736/1/72}

\bibitem[{{Kraus} {et~al.}(2012){Kraus}, {Monnier}, {Che}, {Schaefer},
  {Touhami}, {Gies}, {Aufdenberg}, {Baron}, {Thureau}, {ten Brummelaar},
  {McAlister}, {Turner}, {Sturmann}, \& {Sturmann}}]{Kraus2012}
{Kraus}, S., {Monnier}, J.~D., {Che}, X., {et~al.} 2012, \apj, 744, 19,
  \dodoi{10.1088/0004-637X/744/1/19}

\bibitem[{{Kuhn} {et~al.}(2023){Kuhn}, {Benjamin}, {Ishida}, {de Souza},
  {Peloton}, \& {Veneri}}]{Kuhn2023}
{Kuhn}, M.~A., {Benjamin}, R.~A., {Ishida}, E. E.~O., {et~al.} 2023, Research
  Notes of the American Astronomical Society, 7, 57,
  \dodoi{10.3847/2515-5172/acc4c9}

\bibitem[{{Kuhn} {et~al.}(2021){Kuhn}, {de Souza}, {Krone-Martins},
  {Castro-Ginard}, {Ishida}, {Povich}, {Hillenbrand}, \& {COIN
  Collaboration}}]{Kuhn2021}
{Kuhn}, M.~A., {de Souza}, R.~S., {Krone-Martins}, A., {et~al.} 2021, \apjs,
  254, 33, \dodoi{10.3847/1538-4365/abe465}

\bibitem[{{Kuhn} {et~al.}(2024){Kuhn}, {Hillenbrand}, {Connelley}, {Rich},
  {Staels}, {Carvalho}, {Lucas}, {Fremling}, {Karambelkar}, {Lee}, {Ahumada},
  {Ishida}, {De}, {de Souza}, \& {Kasliwal}}]{kuhn2024}
{Kuhn}, M.~A., {Hillenbrand}, L.~A., {Connelley}, M.~S., {et~al.} 2024, \mnras,
  529, 2630, \dodoi{10.1093/mnras/stae205}

\bibitem[{{Kwan} \& {Fischer}(2011)}]{KwanFischer2011}
{Kwan}, J., \& {Fischer}, W. 2011, \mnras, 411, 2383,
  \dodoi{10.1111/j.1365-2966.2010.17863.x}

\bibitem[{{Lorenzetti} {et~al.}(2009){Lorenzetti}, {Larionov}, {Giannini},
  {Arkharov}, {Antoniucci}, {Nisini}, \& {Di Paola}}]{Lorenzetti2009}
{Lorenzetti}, D., {Larionov}, V.~M., {Giannini}, T., {et~al.} 2009, \apj, 693,
  1056, \dodoi{10.1088/0004-637X/693/2/1056}

\bibitem[{{Mainzer} {et~al.}(2011){Mainzer}, {Bauer}, {Grav}, {Masiero},
  {Cutri}, {Dailey}, {Eisenhardt}, {McMillan}, {Wright}, {Walker}, {Jedicke},
  {Spahr}, {Tholen}, {Alles}, {Beck}, {Brandenburg}, {Conrow}, {Evans},
  {Fowler}, {Jarrett}, {Marsh}, {Masci}, {McCallon}, {Wheelock}, {Wittman},
  {Wyatt}, {DeBaun}, {Elliott}, {Elsbury}, {Gautier}, {Gomillion}, {Leisawitz},
  {Maleszewski}, {Micheli}, \& {Wilkins}}]{Mainzer2011}
{Mainzer}, A., {Bauer}, J., {Grav}, T., {et~al.} 2011, \apj, 731, 53,
  \dodoi{10.1088/0004-637X/731/1/53}

\bibitem[{{Mamajek} {et~al.}(2015){Mamajek}, {Torres}, {Prsa}, {Harmanec},
  {Asplund}, {Bennett}, {Capitaine}, {Christensen-Dalsgaard}, {Depagne},
  {Folkner}, {Haberreiter}, {Hekker}, {Hilton}, {Kostov}, {Kurtz}, {Laskar},
  {Mason}, {Milone}, {Montgomery}, {Richards}, {Schou}, \&
  {Stewart}}]{Mamajek2015}
{Mamajek}, E.~E., {Torres}, G., {Prsa}, A., {et~al.} 2015, arXiv e-prints,
  arXiv:1510.06262.
\newblock \doarXiv{1510.06262}

\bibitem[{{Martin}(1996)}]{Martin1996}
{Martin}, S.~C. 1996, \apj, 470, 537, \dodoi{10.1086/177886}

\bibitem[{{Marton} {et~al.}(2016){Marton}, {T{\'o}th}, {Paladini}, {Kun},
  {Zahorecz}, {McGehee}, \& {Kiss}}]{Marton2016}
{Marton}, G., {T{\'o}th}, L.~V., {Paladini}, R., {et~al.} 2016, \mnras, 458,
  3479, \dodoi{10.1093/mnras/stw398}

\bibitem[{{Masci} {et~al.}(2019){Masci}, {Laher}, {Rusholme}, {Shupe}, {Groom},
  {Surace}, {Jackson}, {Monkewitz}, {Beck}, {Flynn}, {Terek}, {Landry},
  {Hacopians}, {Desai}, {Howell}, {Brooke}, {Imel}, {Wachter}, {Ye}, {Lin},
  {Cenko}, {Cunningham}, {Rebbapragada}, {Bue}, {Miller}, {Mahabal}, {Bellm},
  {Patterson}, {Juri{\'c}}, {Golkhou}, {Ofek}, {Walters}, {Graham}, {Kasliwal},
  {Dekany}, {Kupfer}, {Burdge}, {Cannella}, {Barlow}, {Van Sistine}, {Giomi},
  {Fremling}, {Blagorodnova}, {Levitan}, {Riddle}, {Smith}, {Helou}, {Prince},
  \& {Kulkarni}}]{Masci2019}
{Masci}, F.~J., {Laher}, R.~R., {Rusholme}, B., {et~al.} 2019, \pasp, 131,
  018003, \dodoi{10.1088/1538-3873/aae8ac}

\bibitem[{{Megeath} {et~al.}(2012){Megeath}, {Gutermuth}, {Muzerolle},
  {Kryukova}, {Flaherty}, {Hora}, {Allen}, {Hartmann}, {Myers}, {Pipher},
  {Stauffer}, {Young}, \& {Fazio}}]{Megeath2012}
{Megeath}, S.~T., {Gutermuth}, R., {Muzerolle}, J., {et~al.} 2012, \aj, 144,
  192, \dodoi{10.1088/0004-6256/144/6/192}

\bibitem[{{Meyer} {et~al.}(1997){Meyer}, {Calvet}, \&
  {Hillenbrand}}]{Meyer1997}
{Meyer}, M.~R., {Calvet}, N., \& {Hillenbrand}, L.~A. 1997, \aj, 114, 288,
  \dodoi{10.1086/118474}

\bibitem[{{Muzerolle} {et~al.}(2001){Muzerolle}, {Calvet}, \&
  {Hartmann}}]{Muzerolle2001}
{Muzerolle}, J., {Calvet}, N., \& {Hartmann}, L. 2001, \apj, 550, 944,
  \dodoi{10.1086/319779}

\bibitem[{{Nagy} {et~al.}(2022){Nagy}, {{\'A}brah{\'a}m}, {K{\'o}sp{\'a}l},
  {Park}, {Siwak}, {Cruz-S{\'a}enz de Miera}, {Fiorellino},
  {Garc{\'\i}a-{\'A}lvarez}, {Szab{\'o}}, {Antoniucci}, {Giannini}, {Giunta},
  {Kriskovics}, {Kun}, {Marton}, {Mo{\'o}r}, {Nisini}, {P{\'a}l}, {Szabados},
  {Zieli{\'n}ski}, \& {Wyrzykowski}}]{Nagy2022}
{Nagy}, Z., {{\'A}brah{\'a}m}, P., {K{\'o}sp{\'a}l}, {\'A}., {et~al.} 2022,
  \mnras, 515, 1774, \dodoi{10.1093/mnras/stac1915}

\bibitem[{{Nagy} {et~al.}(2023){Nagy}, {Park}, {{\'A}brah{\'a}m},
  {K{\'o}sp{\'a}l}, {Cruz-S{\'a}enz de Miera}, {Kun}, {Siwak}, {Szab{\'o}},
  {Szil{\'a}gyi}, {Fiorellino}, {Giannini}, {Lee}, {Lee}, {Marton}, {Szabados},
  {Vitali}, {Andrzejewski}, {Gromadzki}, {Hodgkin}, {Jab{\l}o{\'n}ska},
  {Mendez}, {Merc}, {Michniewicz}, {Miko{\l}ajczyk}, {Pylypenko}, {Ratajczak},
  {Wyrzykowski}, {Zejmo}, \& {Zieli{\'n}ski}}]{Nagy2023}
{Nagy}, Z., {Park}, S., {{\'A}brah{\'a}m}, P., {et~al.} 2023, \mnras, 524,
  3344, \dodoi{10.1093/mnras/stad2019}

\bibitem[{{Nisini} {et~al.}(2004){Nisini}, {Antoniucci}, \&
  {Giannini}}]{Nisini2004}
{Nisini}, B., {Antoniucci}, S., \& {Giannini}, T. 2004, \aap, 421, 187,
  \dodoi{10.1051/0004-6361:20034386}

\bibitem[{{Park} {et~al.}(2021){Park}, {K{\'o}sp{\'a}l}, {Cruz-S{\'a}enz de
  Miera}, {Siwak}, {Dr{\'o}{\.z}d{\.z}}, {Ign{\'a}cz}, {Jaffe},
  {K{\"o}nyves-T{\'o}th}, {Kriskovics}, {Lee}, {Lee}, {Mace}, {Og{\l}oza},
  {P{\'a}l}, {Potter}, {Szab{\'o}}, {Sefako}, \& {Worters}}]{Park2021}
{Park}, S., {K{\'o}sp{\'a}l}, {\'A}., {Cruz-S{\'a}enz de Miera}, F., {et~al.}
  2021, \apj, 923, 171, \dodoi{10.3847/1538-4357/ac29c4}

\bibitem[{{Park} {et~al.}(2022){Park}, {K{\'o}sp{\'a}l}, {{\'A}brah{\'a}m},
  {Cruz-S{\'a}enz de Miera}, {Fiorellino}, {Siwak}, {Nagy}, {Giannini},
  {Carini}, {Szab{\'o}}, {Lee}, {Lee}, {Vitali}, {Kun}, {Cseh}, {Krezinger},
  {Kriskovics}, {Ordasi}, {P{\'a}l}, {Szak{\'a}ts}, {Vida}, \&
  {Vink{\'o}}}]{Park2022}
{Park}, S., {K{\'o}sp{\'a}l}, {\'A}., {{\'A}brah{\'a}m}, P., {et~al.} 2022,
  \apj, 941, 165, \dodoi{10.3847/1538-4357/aca01e}

\bibitem[{{Pecaut} \& {Mamajek}(2013)}]{PecautMamajek2013}
{Pecaut}, M.~J., \& {Mamajek}, E.~E. 2013, \apjs, 208, 9,
  \dodoi{10.1088/0067-0049/208/1/9}

\bibitem[{{Podio} {et~al.}(2008){Podio}, {Garcia}, {Bacciotti}, {Antoniucci},
  {Nisini}, {Dougados}, \& {Takami}}]{Podio2008}
{Podio}, L., {Garcia}, P.~J.~V., {Bacciotti}, F., {et~al.} 2008, \aap, 480,
  421, \dodoi{10.1051/0004-6361:20078694}

\bibitem[{{Rayner} {et~al.}(2003){Rayner}, {Toomey}, {Onaka}, {Denault},
  {Stahlberger}, {Vacca}, {Cushing}, \& {Wang}}]{Rayner2003}
{Rayner}, J.~T., {Toomey}, D.~W., {Onaka}, P.~M., {et~al.} 2003, \pasp, 115,
  362, \dodoi{10.1086/367745}

\bibitem[{{Shappee} {et~al.}(2014){Shappee}, {Prieto}, {Grupe}, {Kochanek},
  {Stanek}, {De Rosa}, {Mathur}, {Zu}, {Peterson}, {Pogge}, {Komossa}, {Im},
  {Jencson}, {Holoien}, {Basu}, {Beacom}, {Szczygie{\l}}, {Brimacombe},
  {Adams}, {Campillay}, {Choi}, {Contreras}, {Dietrich}, {Dubberley},
  {Elphick}, {Foale}, {Giustini}, {Gonzalez}, {Hawkins}, {Howell}, {Hsiao},
  {Koss}, {Leighly}, {Morrell}, {Mudd}, {Mullins}, {Nugent}, {Parrent},
  {Phillips}, {Pojmanski}, {Rosing}, {Ross}, {Sand}, {Terndrup}, {Valenti},
  {Walker}, \& {Yoon}}]{shappee2014}
{Shappee}, B.~J., {Prieto}, J.~L., {Grupe}, D., {et~al.} 2014, \apj, 788, 48,
  \dodoi{10.1088/0004-637X/788/1/48}

\bibitem[{{Shingles} {et~al.}(2021){Shingles}, {Smith}, {Young}, {Smartt},
  {Tonry}, {Denneau}, {Heinze}, {Weiland}, {Flewelling}, {Stalder},
  {Clocchiatti}, {F{\"o}rster}, {Pignata}, {Rest}, {Anderson}, {Stubbs}, \&
  {Erasmus}}]{Shingles2021}
{Shingles}, L., {Smith}, K.~W., {Young}, D.~R., {et~al.} 2021, Transient Name
  Server AstroNote, 7, 1

\bibitem[{{Siess} {et~al.}(2000){Siess}, {Dufour}, \& {Forestini}}]{Siess2000}
{Siess}, L., {Dufour}, E., \& {Forestini}, M. 2000, \aap, 358, 593.
\newblock \doarXiv{astro-ph/0003477}

\bibitem[{{Siwak} {et~al.}(2023){Siwak}, {Hillenbrand}, {K{\'o}sp{\'a}l},
  {{\'A}brah{\'a}m}, {Giannini}, {De}, {Mo{\'o}r}, {Szil{\'a}gyi},
  {Jan{\'\i}k}, {Koen}, {Park}, {Nagy}, {Cruz-S{\'a}enz de Miera},
  {Fiorellino}, {Marton}, {Kun}, {Lucas}, {Udalski}, \&
  {Szab{\'o}}}]{Siwak2023}
{Siwak}, M., {Hillenbrand}, L.~A., {K{\'o}sp{\'a}l}, {\'A}., {et~al.} 2023,
  \mnras, 524, 5548, \dodoi{10.1093/mnras/stad2135}

\bibitem[{{Smith} {et~al.}(2020){Smith}, {Smartt}, {Young}, {Tonry}, {Denneau},
  {Flewelling}, {Heinze}, {Weiland}, {Stalder}, {Rest}, {Stubbs}, {Anderson},
  {Chen}, {Clark}, {Do}, {F{\"o}rster}, {Fulton}, {Gillanders}, {McBrien},
  {O'Neill}, {Srivastav}, \& {Wright}}]{Smith2020}
{Smith}, K.~W., {Smartt}, S.~J., {Young}, D.~R., {et~al.} 2020, \pasp, 132,
  085002, \dodoi{10.1088/1538-3873/ab936e}

\bibitem[{{Szegedi-Elek} {et~al.}(2020){Szegedi-Elek}, {{\'A}brah{\'a}m},
  {Wyrzykowski}, {Kun}, {K{\'o}sp{\'a}l}, {Chen}, {Marton}, {Mo{\'o}r}, {Kiss},
  {P{\'a}l}, {Szabados}, {Varga}, {Varga-Vereb{\'e}lyi}, {Andreas}, {Bachelet},
  {Bischoff}, {B{\'o}di}, {Breedt}, {Burgaz}, {Butterley}, {Carrasco},
  {{\v{C}}epas}, {Damljanovic}, {Gezer}, {Godunova}, {Gromadzki}, {Gurgul},
  {Hardy}, {Hildebrandt}, {Hoffmann}, {Hundertmark}, {Ihanec}, {Janulis},
  {Kalup}, {Kaczmarek}, {K{\"o}nyves-T{\'o}th}, {Krezinger}, {Kruszy{\'n}ska},
  {Littlefair}, {Maskoli{\={u}}nas}, {M{\'e}sz{\'a}ros}, {Miko{\l}ajczyk},
  {Mugrauer}, {Netzel}, {Ordasi}, {Pak{\v{s}}tien{\.{e}}}, {Rybicki},
  {S{\'a}rneczky}, {Seli}, {Simon}, {{\v{S}}i{\v{s}}kauskait{\.{e}}},
  {S{\'o}dor}, {Sokolovsky}, {Stenglein}, {Street}, {Szak{\'a}ts}, {Tomasella},
  {Tsapras}, {Vida}, {Zdanavi{\v{c}}ius}, {Zieli{\'n}ski}, {Zieli{\'n}ski}, \&
  {Zi{\'o}{\l}kowska}}]{SzegediElek2020}
{Szegedi-Elek}, E., {{\'A}brah{\'a}m}, P., {Wyrzykowski}, {\L}., {et~al.} 2020,
  \apj, 899, 130, \dodoi{10.3847/1538-4357/aba129}

\bibitem[{{Szil{\'a}gyi} {et~al.}(2021){Szil{\'a}gyi}, {Kun}, \&
  {{\'A}brah{\'a}m}}]{Szilagyi2021}
{Szil{\'a}gyi}, M., {Kun}, M., \& {{\'A}brah{\'a}m}, P. 2021, \mnras, 505,
  5164, \dodoi{10.1093/mnras/stab1496}

\bibitem[{{Tody}(1986)}]{Tody1986}
{Tody}, D. 1986, in Society of Photo-Optical Instrumentation Engineers (SPIE)
  Conference Series, Vol. 627, Instrumentation in astronomy VI, ed. D.~L.
  {Crawford}, 733, \dodoi{10.1117/12.968154}

\bibitem[{{Tonry} {et~al.}(2018){Tonry}, {Denneau}, {Heinze}, {Stalder},
  {Smith}, {Smartt}, {Stubbs}, {Weiland}, \& {Rest}}]{Tonry2018}
{Tonry}, J.~L., {Denneau}, L., {Heinze}, A.~N., {et~al.} 2018, \pasp, 130,
  064505, \dodoi{10.1088/1538-3873/aabadf}

\bibitem[{{Vacca} {et~al.}(2003){Vacca}, {Cushing}, \& {Rayner}}]{Vacca2003}
{Vacca}, W.~D., {Cushing}, M.~C., \& {Rayner}, J.~T. 2003, \pasp, 115, 389,
  \dodoi{10.1086/346193}

\bibitem[{{Vacca} \& {Sandell}(2011)}]{VaccaSandell2011}
{Vacca}, W.~D., \& {Sandell}, G. 2011, \apj, 732, 8,
  \dodoi{10.1088/0004-637X/732/1/8}

\bibitem[{{Whelan} {et~al.}(2014){Whelan}, {Bonito}, {Antoniucci},
  {Alcal{\'a}}, {Giannini}, {Nisini}, {Bacciotti}, {Podio}, {Stelzer}, \&
  {Comer{\'o}n}}]{Whelan2014}
{Whelan}, E.~T., {Bonito}, R., {Antoniucci}, S., {et~al.} 2014, \aap, 565, A80,
  \dodoi{10.1051/0004-6361/201322037}

\bibitem[{{Wright} {et~al.}(2010){Wright}, {Eisenhardt}, {Mainzer}, {Ressler},
  {Cutri}, {Jarrett}, {Kirkpatrick}, {Padgett}, {McMillan}, {Skrutskie},
  {Stanford}, {Cohen}, {Walker}, {Mather}, {Leisawitz}, {Gautier}, {McLean},
  {Benford}, {Lonsdale}, {Blain}, {Mendez}, {Irace}, {Duval}, {Liu}, {Royer},
  {Heinrichsen}, {Howard}, {Shannon}, {Kendall}, {Walsh}, {Larsen}, {Cardon},
  {Schick}, {Schwalm}, {Abid}, {Fabinsky}, {Naes}, \& {Tsai}}]{Wright2010}
{Wright}, E.~L., {Eisenhardt}, P. R.~M., {Mainzer}, A.~K., {et~al.} 2010, \aj,
  140, 1868, \dodoi{10.1088/0004-6256/140/6/1868}

\end{thebibliography}
\bibliographystyle{aasjournal}



\end{document}